\documentclass[12pt,preprint]{aastex6} 
\usepackage{subfigure}
\usepackage{graphicx}
\usepackage{amsmath}

\slugcomment{Draft, revised 1, \today}

\shorttitle{New Suns in the Cosmos III}
\shortauthors{D. B. de Freitas et al.}

\begin{document}

\title{New Suns in the Cosmos III: multifractal signature analysis}

\author{D. B. de Freitas, M. M. F. Nepomuceno, P. R. V. de Moraes Junior, C. E. F. Lopes, M. L. Das Chagas, J. P. Bravo, A. D. Costa, B. L. Canto Martins, J. R. De Medeiros}
\affil{Departamento de F\'{\i}sica,
    Universidade Federal do Rio
    Grande do Norte, 59072-970
    Natal, RN, Brazil}
\and
\author{I. C. Le\~ao}
\affil{European Southern Observatory, Karl-Schwarzschild-Str. 2, 85748 Garching, Germany}

\begin{abstract}
In present paper, we investigate the multifractality signatures in hourly time series extracted from CoRoT spacecraft database. 
Our analysis is intended to highlight the possibility that astrophysical time series can be members of a particular class of complex and dynamic
processes which require several photometric variability diagnostics to characterize their structural and topological properties. To achieve this goal, we search for contributions due to nonlinear temporal correlation and effects caused by heavier tails than the Gaussian distribution, using a detrending moving average algorithm for one-dimensional multifractal signals (MFDMA). We observe that the correlation structure is the main source of multifractality, while heavy-tailed distribution plays a minor role in generating the multifractal effects. Our work also reveals that rotation period of stars is inherently scaled by degree of multifractality. As a result, analyzing the multifractal degree of referred series, we uncover an evolution of multifractality from shorter to larger periods.
\end{abstract}


\section{Introduction}\label{sec:intro}
In different areas of knowledge, many phenomena have been demonstrated to be governed by nonlinear dissipative systems and manifest (multi)fractal and complex structures in one-dimensional signals and images \citep{hurst1965,feder1988,komm1995,ivanov1999,cs2001,Norouzzadeha,defreitas2013b}. In particular, these phenomena exhibit linear or nonlinear behavior depending on how the output energy is related to the input energy of the system \citep{mw1969a,mw1969b,mw1969c,aschwa,a2011}. This transitional process is associated with entities that interact in a complex manner that requires $(i)$ a continuous energy input source and $(ii)$ a nonlinear dissipative system. In stellar physics, the mechanisms of the solar dynamo operation can be supported by a dominance of large or small fluctuations in the total energy emitted by solar indicators, such as flares and sunspots \citep{movahed,sen2007,defreitas2009}.

Recently, \cite{defreitas2013} discussed the statistical and fractal properties of CoRoT time series due to cyclic behavior attributed to magnetic activity on the stellar photosphere, particularly rotational modulation \citep{baglin2006}. The importance of astrophysical time series analyses, which typically exhibit non-stationary and nonlinear dynamics, has been recognized in the area of complex systems analysis and self-organized criticality (SOC) \citep{watari,aschwa}. Several features of these approaches have been adopted to detect the spatiotemporal dynamical behavior of astrophysical phenomena. An important approach based on measuring the smoothness of a fractal time series and thereby characterizing signals is the traditional rescale range analysis ($R/S$) method based on a power-law with slope $H$, i.e., the Hurst exponent \citep{hurst1951,hurst1965}. $R/S$ analysis is a statistical technique used for detecting persistence, randomness, or mean reversion in time series through the exponent $H$ statistically known as an index of long-memory process. Using this method, \cite{defreitas2013} demonstrated that different centers of rotation due to modulation affect the slope $H$ on different timescales, corresponding to sub-harmonics of rotation period of each star. These centers of rotation can be associated to chaotic motions temporally aperiodic (neither random, nor periodic) due to a fractal hierarchic structure \citep{sps2009,defreitas2013}. 

The variation in the Hurst exponent $H$ in different time windows can be used to quantify dynamic changes in the features of a time series. As an example, lower $H$ values characterize the portions of the signal with lower complexity. In contrast, higher values reveal strong interactions in the time series dynamics. This behavior clearly complements a wide range of methods that are routinely used in astrophysics, such as Lomb-Scargle periodograms, autocorrelation functions and Gaussian processes \citep{lomb1976,scargle1982}. Similarly, different levels of complexity can be associated with an anti-persistency/persistency (i.e., whether fractional Brownian motion is obeyed) processes extracted by a fractal spectral analysis in terms of the local Hurst exponent. As noted by \cite{defreitas2013}, the exponent $H$ can indicate a quantification of variability in relatively brief and noisy time series. 

In this sense, astrophysical time series can be classified as members of a class of physical ensembles that involve long-range interactions (complex fluctuations), long-range microscopic memories (e.g., non-Markovian stochastic processes), or multifractal structures. In this context, time-domain astrophysics cannot be treated using traditional statistical analysis methods for a wide range of problems \citep{feigelson}. Several sources of multifractality are not captured by conventional measures, such as Fourier and wavelet analysis \citep{bravo}. For instance, a rotational signature has a scale-invariant structure, i.e, it do not change if scales of length of the time windows are multiplied by a factor. Mathematically, a time series $x(t)$ is scale-free when $x(ct)=c^{H}x(t)$, i.e., it does not obey the homogeneity property and it is described by a power law rather than a Gaussian distribution, which is typical in monofractal systems \citep{a2011,ihlen}. Signals with structures that are independent of time and space are known as monofractals and are defined by a single power law exponent, i.e., one unique value for $H$. Nonetheless, spatial and temporal variations in scale-invariant structures are most common in astrophysical scenarios. In this case, they are best defined by a $H$ spectrum, i.e., the Holder spectrum \citep{defreitas2013}. 

Some authors \citep[e.g.,][]{sps2009,defreitas2013} have shown that the variations in stellar magnetic activity present a global value of $H$ greater than 0.5, indicating long-term memory in the time series \citep{Kantelhardt}. However, a multifractal time series not only exhibits long-range correlations over different time scales but can also result from heavy tails in the probability distributions even if the data have no memory \citep{gp}. In finite datasets, large fluctuations cannot be detected, only small fluctuations. In other words, the multifractal properties of shorter time series reveal a wide variety of fluctuations at different scales, whereas longer time series are corrupted by various effects, such as noise, short-term memory or periodic signals \citep{gp}. 

\cite{defreitas2013} observed fractality traces in the time series of the Sun in its active and quiet phases and in a sample of 14 CoRoT stars with sub- and super-solar rotational periods and 3 stars with period near the solar value \citep{lanza2003}. It is worth noting that our selected stars present values of log$T_{eff}$ (effective temperature) between $\sim$ 3.62 and $\sim$ 4.39 and log $g$ (effective gravity) from 2.8 to 4.6 \citep{demedeiros2013}. These researchers computed the global Hurst exponent for these stars and found a clear correlation between the Hurst exponent and rotational period. This result reveals that the global Hurst exponent can be considered as a powerful classifier for noisy semi-sinusoidal time series. However, the $R/S$ analysis is considered under the hypothesis that the working sample is monofractal, i.e., characterized by a global singularity exponent. In addition, local scaling exponents can be revealed considering a priori that the sample is intrinsically more complex and dynamic than a monofractal. There are a vast number of methods that have been developed to investigate the behavior and properties of monofractal and multifractal systems. Among them, we can cite classic rescaled range analysis ($R/S$) \citep{hurst1951,hurst1965} and the detrended fluctuation analysis (DFA) \citep{taqqu} and detrended moving average (DMA) algorithm \citep{alessio}, all of which have been appropriated for fractal time series.  The wavelet-based transform module maxima (WTMM) \citep{muzy1991,muzy1994}, multifractal detrended fluctuation analysis (MFDFA) \citep{Kantelhardt} and multifractal detrended moving average (MFDMA) \citep{gu2010} are commonly used for multifractal time series. Independent of the (multi)fractal analysis method applied, the multifractal spectrum is one of most important products used to study complex time series \citep[cf.][]{gu2010}. Previous studies have noted that MFDMA is slightly better than WTMM and MFDFA for the performance and multifractal characterization of time series data \citep{ruan,zhou}.

The main purpose of the present work is to examine the correlation between the multifractality signatures and stellar variability in time series of less than 150 days, a phenomenon that has been investigated in a pioneering study published by \cite{defreitas2013}. Indeed, we will investigate the behavior of the variability in the same CoRoT sample analyzed by these authors. In this context, we will compare the results of the multifractal analysis for original, shuffled and phase-randomized data and verify whether the processes are affected by strong correlations and nonlinearity, among other features. 
The remainder of this paper is organized as follows. In the next section, we describe the methods used in our analysis. In section 3, we provide our results and discuss their implications, and in the last section, we present our final remarks. 

\section{Procedures and multifractal analysis}

The main feature of multifractals is that the fractal dimension is not the same on all scales. In case of a one-dimensional signal, the fractal dimension can vary from unity to a dot. This property reveals a variability spectrum on a wide range of temporal scales, associated with quasi- or periodic fluctuations and power-law-like correlations due to astrophysical noise. For our particular case, we have chosen an extended version of the detrended moving average (DMA) algorithm, which consists of a multifractal characterization of nonstationary time series \citep{gu2010}. Whereas the DMA method generates the same scaling properties throughout the entire signal indexed by a single global Hurst exponent, the MFDMA details the signal on a wide Hurst exponent spectrum. Thus, the Hurst exponent defined by the DMA method represents the average fractal structure of the entire signal, thus representing the central tendency of the multifractal spectrum \citep{alessio}. 

Our sample is composed of time series of the Sun (Virgo/SOHO light curves in two regimes: active and quiet) and 14 obtained from CoRoT spacecraft (all presenting rotational modulation) with log$T_{eff}$ varying from 3.6 to 3.8, i.e., from fully convective M stars to the Kraft break \citep{kraft1967}. The time series examined in our work present a regular dynamics due to rotational modulation in a wide interval of frequencies and irregular dynamics due to astrophysical noise, thus making them suitable candidates for multifractal analysis. We applied the one-dimensional multifractal detrending moving average analysis (MFDMA) according to the procedure summarized in \citep{gu2010}. 

In general, a time series is composed by a deterministic function $p(t,\bar{P})$, where the vector $P$ is a Fourier series, and a stochastic term characterized by astrophysical noise (without instrumental bias) given by $\eta(t)$. In this sense, a time series $x(t)$ can be written as
\begin{equation}
\label{eq1}
x(t)=p(t,\bar{P})+\eta(t),
\end{equation}
where $\eta(t)$ represents the colorful noise associated with the $1/f^{\gamma}$-like one  in \cite{cw2009}. The parameter $\gamma$ consists of the color index of the noise and is associated with the global Hurst exponent $H$ by the relation $\gamma=H\pm 1$, as noted by \cite{pg2011} and \cite{defreitas2013}.

In general, all procedures that involve a multifractal background generate three crucial parameters to describe the structural properties of a time series $x(t)$: $(i)$ a $q_{th}$-order fluctuation function $F_{q}(n)$, $(ii)$ the multifractal scaling exponent $\tau(q)$ (i.e., the Renyi exponent), and $(iii)$ the multifractal spectrum $f(\alpha)$. These parameters are obtained according to the following steps:
\begin{itemize}
	\item First, the time series is reconstructed as a sequence of cumulative sums given by
\begin{equation}
\label{eq1}
y(t)=\sum^{t}_{i=1}x(t), \quad t=1,2,...,N,
\end{equation}
where $N$ is the length of the signal.
\end{itemize}

\begin{itemize}
	\item Second, the moving average function $\tilde{y}(t)$ of Eq. (\ref{eq1}) is calculated in a moving window:
	\begin{equation}
\label{eq1a}
\tilde{y}(t)=\frac{1}{n}\sum^{\left\lceil (n-1)(1-\theta)\right\rceil}_{k=-\left\lfloor (n-1)\theta\right\rfloor}y(t-k),
\end{equation}
where $n$ is the window size, $\left\lceil (.)\right\rceil$ is the largest integer not larger than argument $(.)$, $\left\lfloor (.)\right\rfloor$ is the smallest integer not smaller than argument $(.)$,
and $\theta$ is the position index with values between 0 and 1. In present work, $\theta$ was adopted as 0, referring to the backward moving average, i.e., $\tilde{y}(t)$ is calculated over all the past $n-1$ data of the time series \citep[for more details, see][]{gu2010};
\end{itemize}

\begin{itemize}
	\item Third, the trend is removed from the reconstructed time series $y(t)$ using the function $\tilde{y}(t)$ and the residual sequence $\epsilon(t)$ is obtained:
	\begin{equation}
\label{eq2}
\epsilon(i)=y(i)-\tilde{y}(i), \quad n-\left\lfloor (n-1)\theta\right\rfloor\leq i\leq N-\left\lfloor (n-1)\theta\right\rfloor
\end{equation}
where, this residual time series $\epsilon(i)$ is subdivided into $N_{n}$ disjoint segments with the same size $n$ given by $\left\lfloor N/n-1\right\rfloor$. In this sense, the residual sequence $\epsilon(t)$ for each segment is denoted by $\epsilon_{\nu}$, where $\epsilon_{\nu}(i)=\epsilon(l+1)$ for $1\leq i\leq n$ and $l=(\nu -1)n$;
\end{itemize}
\begin{itemize}
	\item Fourth, we calculate the root-mean-square (rms) function $F_{\nu}(n)$ for a segment size $n$,
\begin{equation}
\label{eq3}
F_{\nu}(n)=\left\{\frac{1}{n}\sum^{n}_{i=1}\epsilon^{2}_{\nu}(i)\right\}^{\frac{1}{2}}.
\end{equation}
\end{itemize}
\begin{itemize}
	\item Fifth, the generating function  of $F_{q}(n)$ is determined as
\begin{equation}
\label{eq4}
F_{q}(n)=\left\{\frac{1}{N_{n}}\sum^{N_{n}}_{\nu=1}F^{2}_{\nu}(n)\right\}^{\frac{1}{q}}, 
\end{equation}
for all $q\neq 0$ and for $q=0$; hence,
\begin{equation}
\label{eq4b}
F_{q}(n)=\exp{\left\{\frac{1}{2N_{n}}\sum^{N_{n}}_{\nu=1}\ln [F^{2}_{\nu}(n)]\right\}}, 
\end{equation}
where the scaling behavior of $F_{q}(n)$ follows a power-law type given by $F_{q}(n)\sim n^{h(q)}$ and $h(q)$ represents the local Hurst or Holder exponent.
\end{itemize}
\begin{itemize}
	\item Sixth, knowing $h(q)$, we can easily derive $\tau(q)$ using
\begin{equation}
\label{eq5}
\tau(q)=q h(q)-1;
\end{equation}	
\end{itemize}
\begin{itemize}
	\item Finally, we obtain the Holder exponent $\alpha(q)$ and the multifractal spectrum $f(\alpha)$, which is related to $\tau(q)$ via a Legendre transform:
\begin{equation}
\label{eq6}
\alpha(q)=\frac{d\tau(q)}{dq},
\end{equation}	
where the multifractal spectrum is defined as
\begin{equation}
\label{eq7}
f(\alpha)=q\alpha-\tau(q).
\end{equation}
\end{itemize}

It is important to stress that in present study, the backward MFDMA algorithm ($\theta=0$) gives us a best computation accuracy of the exponent $\tau(q)$, as well as mentioned by \cite{gu2010}.

As noted by \cite{telesca2006}, the multifractal spectrum provides detailed information about the relative importance of several types of fractal exponents present in the signal. To quantitatively characterize the multifractal spectrum, we investigate the behavior of two features extracted from the [$\alpha,f(\alpha)$] plot: the width of $\alpha$, which indicates the degree of multifractality, and the asymmetry in the shape of $\alpha$. The width of $\alpha$ is defined as $\Delta\alpha=\alpha_{max}-\alpha_{min}$; we measured this width by extrapolating the fitted curve to zero, i.e., by determining the roots of a 4th-order polynomial. The parameter $\Delta\alpha$ can also be understood as a measure of the multifractal diversity or complexity of signal. For instance, larger values of $\Delta\alpha$ indicate a richer and more complex structure, whereas smaller values indicate the monofractal limit. The asymmetry in the shape of $\alpha$ is defined as $A=\frac{\alpha_{max}-\alpha_{0}}{\alpha_{0}-\alpha_{min}}$, where $\alpha_{0}$ is the value of $\alpha$ when $f(\alpha)$ assumes its maximum value. According to the latter relation, the asymmetry $A$ presents three shapes: asymmetry to the right-skewed ($A>1$), left-skewed ($0<A<1$) or symmetric ($A=1$). As noted by \cite{sen2007} and \cite{ihlen}, an asymmetry to a long left tail indicates a multifractal structure that is insensitive to local fluctuations with small magnitudes, whereas a long right tail refers to a structure that is insensitive to local fluctuations with large magnitudes. In the stellar rotation scenario, these fluctuations can be associated with semi-sinusoidal signatures and background noise \citep{demedeiros2013,defreitas2013}. 

In addition to the original data treatment, we apply the MFDMA in two other procedures. These two methods, \textit{shuffling} and \textit{phase randomization}, are used to verify the origin of the multifractality. Shuffling a time series destroys the long-range temporal correlation for small and large fluctuations and preserves the distribution of the data. In other words, the distribution function of the original data remains the same but without memory, i.e., $\alpha_{0}$ is shifted to $\sim 0.5$. The origin of multifractality can also be due to fat-tailed probability distributions present in the original time series. For this case, the non-Gaussian effects can be weakened by creating phase-randomized surrogates. In this context, the procedure preserves the amplitudes of the Fourier transform and the linear properties of the original series but randomizes the Fourier phases while eliminating nonlinear effects \citep{Norouzzadeha}. These procedures are used to study the degree of complexity of time series to distinguish different sources of multifractality in the time series.

Each source can be investigated by using the Holder exponent $h(q)$. Thus, if the shuffled signal only presents long-range correlations, we should find that $h_{shuf}(q)=0.5$. However, if the source of multifractality is due to a heavy-tailed distributions obtained by the surrogate method, the values of $h_{sur}(q)$ will be independent of $q$. An alternative method to assess the behavior of multifractality is to compare the multifractal scaling exponent $\tau(q)$ for the original, shuffled and surrogate data. Differences between these two scaling exponents and the original exponent reveal the presence of long-range correlations and/or heavy-tailed distributions. Using Eq. (\ref{eq5}), this comparison can be shown in a plot, which presents the following relations:
\begin{equation}
\label{eq8}
\tau(q)-\tau_{shuf}(q)=q[h(q)-h_{shuf}(q)]=qh_{corr}(q)
\end{equation}
and
\begin{equation}
\label{eq9}
\tau(q)-\tau_{sur}(q)=q[h(q)-h_{sur}(q)]=qh_{tail}(q),
\end{equation}
where we should find that $h(q)=h_{shuf}(q)$ and thus that $h_{corr}(q)=0$ if the multifractality is due solely to the heaviness of the distribution. The multifractality indicates the presence of long-range correlations if $h_{corr}(q)\neq 0$. However, if the source is due solely to correlations, we will find that $h_{shuf}(q)=0.5$. If both sources are present, $h_{corr}(q)$ and $h_{tail}(q)$ will depend on $q$.

\section{Results and Discussion}\label{sec:results}

The results described by \cite{defreitas2013} suggest a correlation between the Hurst exponent and rotation period (including the Sun), but the source of this correlation was not discussed. In the present paper, we revisit the same sample using a multifractal approach that highlights different features that occur on small and large time scales, whereas the approach used by \cite{defreitas2013} restricts the analysis to global behavior characterized by a single ``mean'' exponent. All original data present in our work are associated with a nonlinear $\tau(q)$ plot indicated by a crossover at $q=0$ (see Figs. \ref{fig1} to \ref{fig4}, left-bottom panels). As reported by \cite{movahed}, the presence of a crossover in $[q-\tau(q)]-plot$, i.e., different slopes for $q<0$ and $q>0$, emphasizes that the multifractality in the time series is extremely strong.

The top panels of Figs. \ref{fig1} through \ref{fig4} illustrate the multifractal fluctuation function $F_{q}(n)$ obtained from the MFDMA for an extract of our stellar sample (including the Sun), where each curve corresponds to different fixed values of $q=-5,...,0,...,5$ in steps of 0.2. The values for the scale parameter $n$ were varied from 10 to $N/10$ \citep{cs2001,gu2010}. In the top-left figures, the top, middle and bottom bold lines correspond to $q=5$, $q=0$ and $q=-5$, respectively; the black lines represent the shuffled data, and the red lines indicate the original data. The different slopes $h(q)$ of each curve indicate that the small and large fluctuations scale differently. The multifractal scaling exponent $\tau(q)$ of the original, shuffled and surrogate time series are compared in the bottom-left panel. The multifractal spectra $f(\alpha)$ of the original, shuffled and surrogate time series are shown in the bottom-right panel, in which the solid curves represent the best 4th-order fit. 

As shown by figures of the active Sun and the best candidate (CoRoT ID 105693572) for a new Sun (figs. \ref{fig1} (left panel) and \ref{fig2} (right panel), respectively), the stars differ from the perspective of a multifractal analysis. As quoted by \cite{defreitas2013}, the ``New Suns'' candidates were selected by the best values for the triplet (Hurst exponent, effective temperature and rotation period), where the star CoRoT ID 105693572 is the best candidate by presenting the values of this triplet most similar to the Sun. In general, some stellar indicators, such as flares and smallest spots, are not detected by instruments because the flux from a light source is proportional to the inverse square of the distance, limiting the instrumental sensitivity of the detectors \citep[for more details, see][]{a2011}. Furthermore, this behavior can also be influenced by the cycle phase of the star. 

\subsection{Results based on resampled data}

As mentioned the previous Section, two different kinds of multifractality can be identified in a time series. One of them destroys the long range correlation by the shuffling procedure, where if the multifractality only is due to this type of correlation, we should find $h_{shuf}(q)=0.5$. These resampled data are shown by green cycles in Figs. \ref{fig1} through \ref{fig4}. As illustrated in Figs. \ref{fig1} and \ref{fig2}, the active Sun is more ``polluted'' by correlated noise on different scales than the \textbf{referred} star CoRoT ID 105693572 using the values of $\alpha_{0}$ of the multifractal spectrum from the shuffled time series as the distinguishing parameter. As discussed in the next section, a white-noise-like time series (shuffled one) has a time-independent structure with $\alpha_{0}\sim 0.5$, as shown by the vast majority of our sample. In contrast, both the active and quiet Sun present values of $\alpha_{0}\sim 1$, indicating that the noise is $1/f^{\gamma}$-like, i.e., a time-correlated noise \citep{cw2009,a2011}.

After shuffling, the time series exhibit an even smaller degree of multifractality than the original one for all stars, except for the quiet Sun and CoRoT ID 105945509. This result can be emphasized by the multifractal spectra, for which $\Delta\alpha-\Delta\alpha_{shuf}>0$. Therefore, as described below, this analysis is insufficient for explaining the source of this behavior. In this case, the behavior of $\tau(q)$ must be verified using the relations \ref{eq8} and \ref{eq9}. As illustrated by Figs. \ref{fig6} and \ref{fig7}, there is a dependence of \textbf{$h_{corr}(q)$} on $q$, which is clearest for $q<0$, where a long right tail occurs, as shown in the lower panels of Figs. \ref{fig1} through \ref{fig4}. 

\subsection{Results based on phase randomized data}

As the shuffling procedure does not remove the multifractality due to a heaviness of distribution, we apply the surrogate method to clarify the contribution due to the phase randomization of discrete Fourier transform coefficients of time series. In contrast to the results found in previous subsection, for $q>0$ the dependence on $q$ can be neglected for the surrogate data in all stellar samples, including the Sun, as shown in Figs. \ref{fig6} and \ref{fig7}. In fact, the absolute values of $h_{corr}(q)$ in both domains ($q<0$ and $q>0$) are typically greater than $h_{tail}(q)$, thus implying that the multifractality due to heaviness is weaker than the multifractality due to long-range correlations. According to the central limit theorem, the superposition of many small and independent effects with finite statistical moments leads to a Gaussian distribution. This means that the small deviations found between the surrogate and original data reduce to the typical condition of the random variables of a system being independent with an approximately linear behavior \citep{Hilhorst}. The results suggest that our sample does not contain important phase correlations that are canceled in the surrogate time series by randomization of the Fourier phases. Thus, the observed multifractality is related to the long-range correlation generated by rotational modulation, except for the quiet Sun. 

\subsection{The degree of multifractality and asymmetry}

Another class of results is associated with the correlation between the stellar parameters and the parameters extracted from our method. The top panel of Fig. \ref{fig5} provides the values for the degree of multifractality ($\Delta\alpha$) based on the multifractal analysis as a function of the rotation period. The solid line denotes the linear regression obtained through a similar adjustment as derived by \cite{defreitas2013}. In this context, the period is related to not only a single exponent but also the multifractal diversity spectrum. 
In contrast, the $\Delta\alpha$--$A$ correlation indicates that different domains can be separated. The values of the parameters extracted from multifractal analysis are summarized in Table \ref{tab1}. This result demonstrates that the transition region between left and right asymmetry, indicated by the horizontal line (see bottom panel in Fig. \ref{fig5}), can serve as a dividing line to distinguish stars in the active and quiet phases. In future works, this procedure can be used as a method for discriminating ``blind'' datasets. Besides, this figure reveals the lack of a clear correlation between the asymmetry parameter and the degree of multifractality.

Finally, in contrast to \cite{defreitas2013}, these behaviors demonstrate that a single exponent is not sufficient to describe the complexity of a stellar time series. It is important to emphasize that there are several alternative methods to deal with the complexity of time series, including time-domain,  wavelets and spectrum-based techniques. In general, these methods are based into three main categories, namely \textit{fractality}, \textit{nonlinear dynamics} and \textit{entropy}, which can broader the range of procedures and tools for handling the problems mentioned in present work. We focus on fractality methods by analysing the geometric properties of a set of data. A literature review of most important methods for analysing time series can be found in \cite{tang}.

\subsection{Interpretation of the physical origin}

We verified for possible correlations between our multifractal indexes ($\Delta\alpha$ and $A$) and effective temperature and surface gravity extracted by \cite{sarro2013}. This analysis indicates that these basic stellar properties are weakly correlated with multifractal indexes, exhibiting absolute values of the Spearman's correlation coefficient $r<0.1$ \citep{press}. Likewise, we also investigate whether this correlation could be caused by another physical source linked to star's brightness variations. Following \cite{karoff}, there are, at least, two approaches of the physical processes responsible for distinct features in solar frequency spectrum. Generally speaking, the manifestation of the different physical processes in time series as a function of frequency, such as granulation, faculae or oscillation, is prone to the cadence of data as reported by \cite{karoff} (see Table 2, here adopted).

In general, as pointed by \cite{moro}, the granular lifetime distributions follow a decaying exponential law, indicating that the signature due to granulation occurs at frequencies higher than 1000$\mu$Hz ($\leq$ 17 min). In agreement with the long-cadence data ($\sim$ 1h) adopted here, this signature can be neglected in our study. At the same time, there is another type of stellar granulation that appears at lower frequencies denoted flicker noise. As reported by \cite{bastien}, flickers are revealed on timescales shorter than 8 hours (in particular, $F_{8}$) and are widely used as a proxy for the surface gravity. In fact, in the present literature, flicker is considered only calibrated to surface gravity. Our results show that there is no evidence of a correlation between the degree of multifractality and the surface gravity and, therefore, we could infer that the effects due to flicker noise can be neglected. However, as shown in Fig. \ref{fig8}, a more depth analysis reveals that $\Delta\alpha$ and $F_{8}$ are correlated with Spearman's coefficient 
of -0.48. The present study applies
the same procedure used by \cite{bastien} and \cite{kipping} to compute the variability in photometric brightness due to $F_{8}$. Thus, our multifractal analysis suggests that the physical origin of the correlation between the rotation period and the degree of multifractality can be due to rotational modulation from long-lived spots \citep{radick,lanza2004}. On the other hand, our results point out that the flicker is a very good candidate for explaining the changes in the multifractal index.

In addition, the CoRoT time series tipically have a cadence shorter than a dozen of minutes \citep{defreitas2013}, allowing us to measure different kind of noisy sources, in general, the photon noise \citep{mathur}. However, a complete investigation about the different indidividual contribution of these sources is out of scope of present paper.

\section{Final remarks}
The multifractal properties of our stellar sample have been investigated in this work through the MFDMA method, which was developed to quantify the long-range correlations of non-stationary time series by using a detrending moving average analysis. Applying this method, we demonstrated that hourly time series of a CoRoT stellar sample are characterized by strong long-range correlations due to rotational modulation. This result suggests that our sample exhibits the features that can be invoked in terms of multifractality approaches. The main insight and contribution of our work is to estimate the different levels of multifractality present according to fluctuations arising from stellar variability. 

\begin{figure*}
\begin{center}
\subfigure{\includegraphics[width=0.48\textwidth]{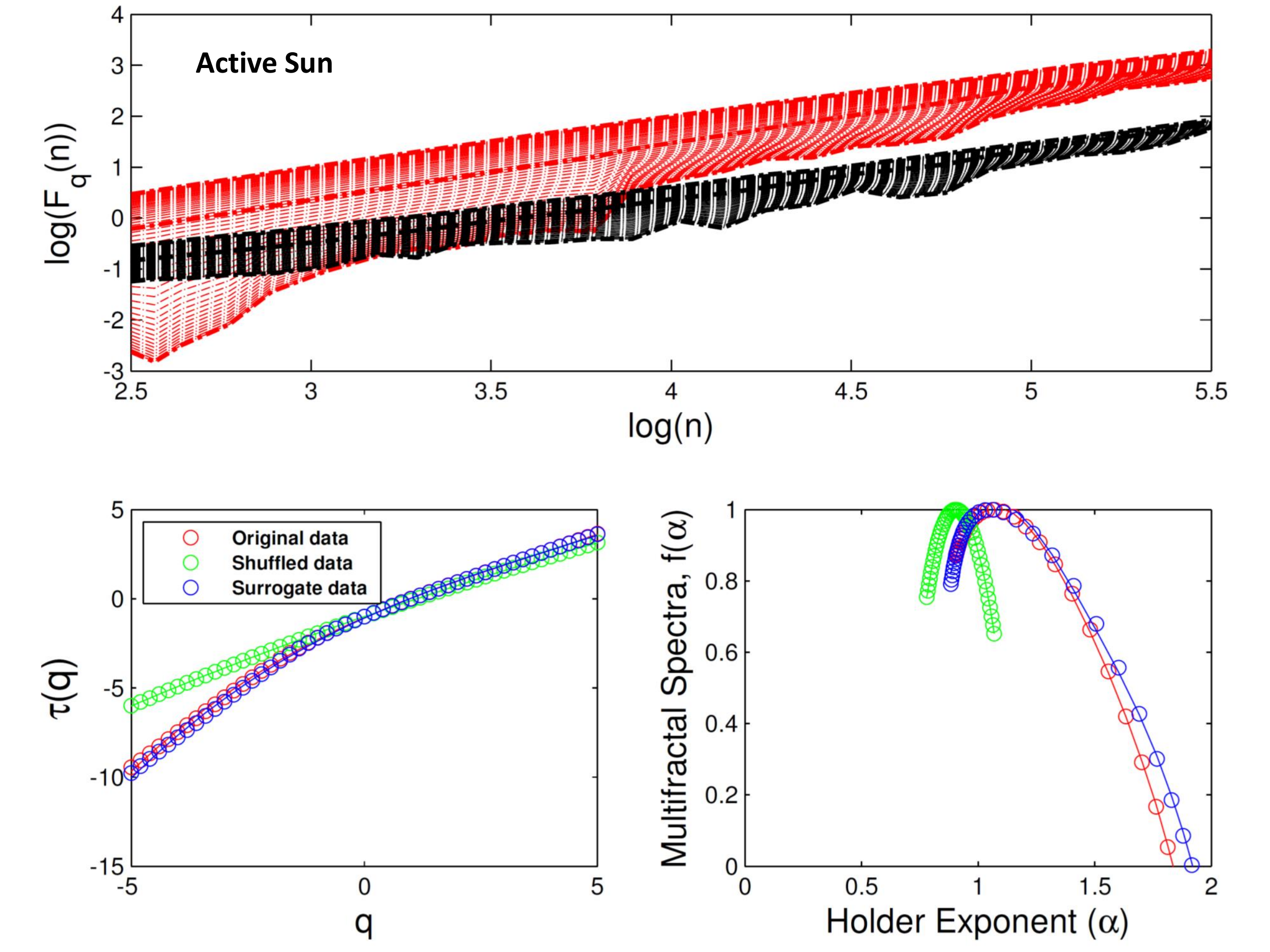}}
\subfigure{\includegraphics[width=0.48\textwidth]{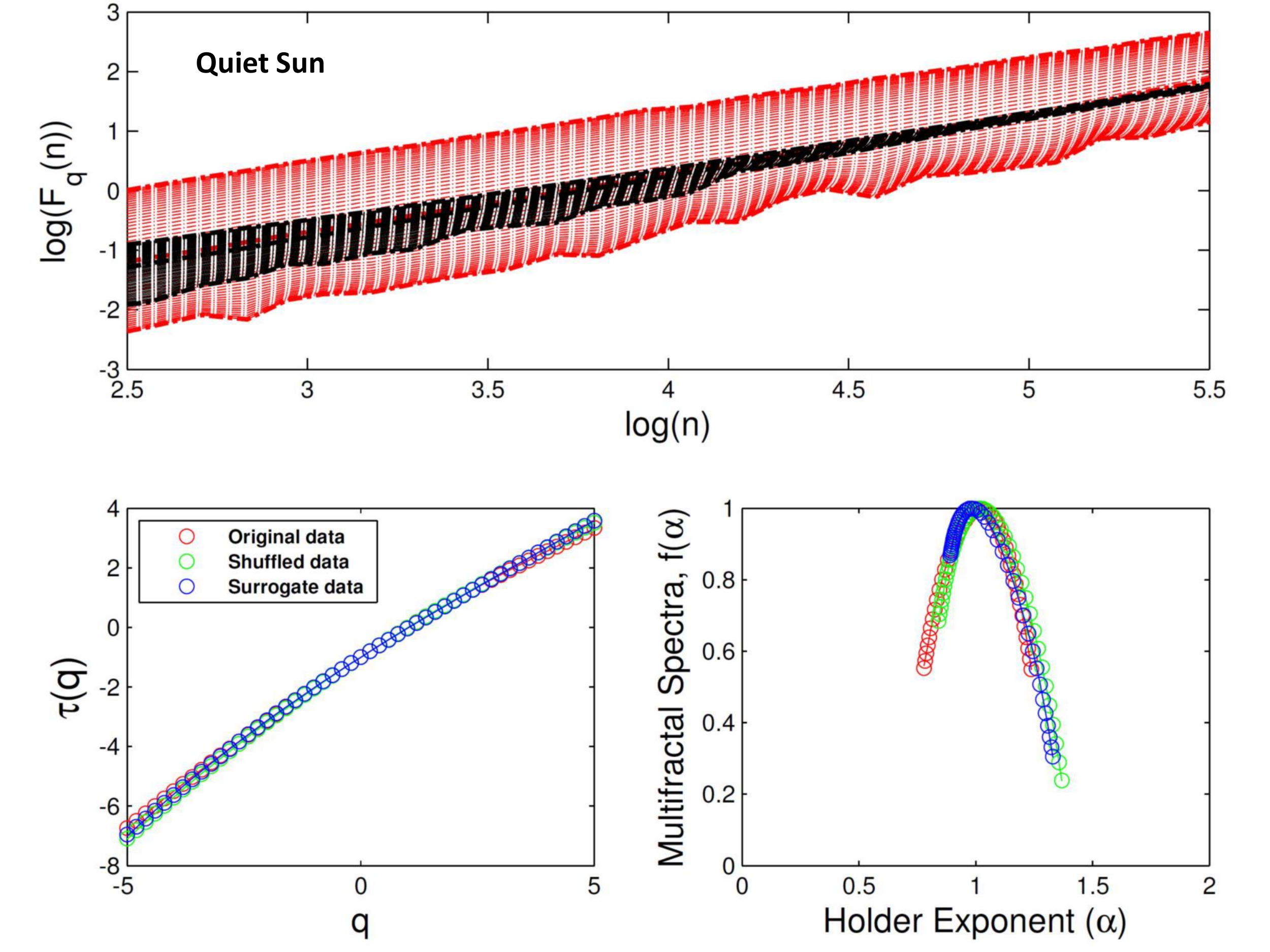}}
\end{center}
\caption{The multifractal fluctuation function $F_{q}(n)$ obtained from MFDMA for active (left panels) and quiet (right panels) Sun. Each curve corresponds
to different fixed values of $q=-5,...,5$ from top to bottom, while top bold, middle bold lines and bottom bold line correspond to $q=5$, $q=0$ and $q=-5$, respectively, where black lines are shuffled data and red lines are original data ($top$). Comparison of the multifractal scaling exponent $\tau(q)$ of the original (red), shuffled (green) and surrogate (blue) time series ($left-bottom$). Multifractal spectrum $f(\alpha)$ of the the original (red), shuffled (green) and surrogate (blue) time series. The solid curves represent the best 4th order fit ($right-bottom$). 
}
\label{fig1}
\end{figure*}

\begin{figure*}
\begin{center}
\subfigure{\includegraphics[width=0.48\textwidth]{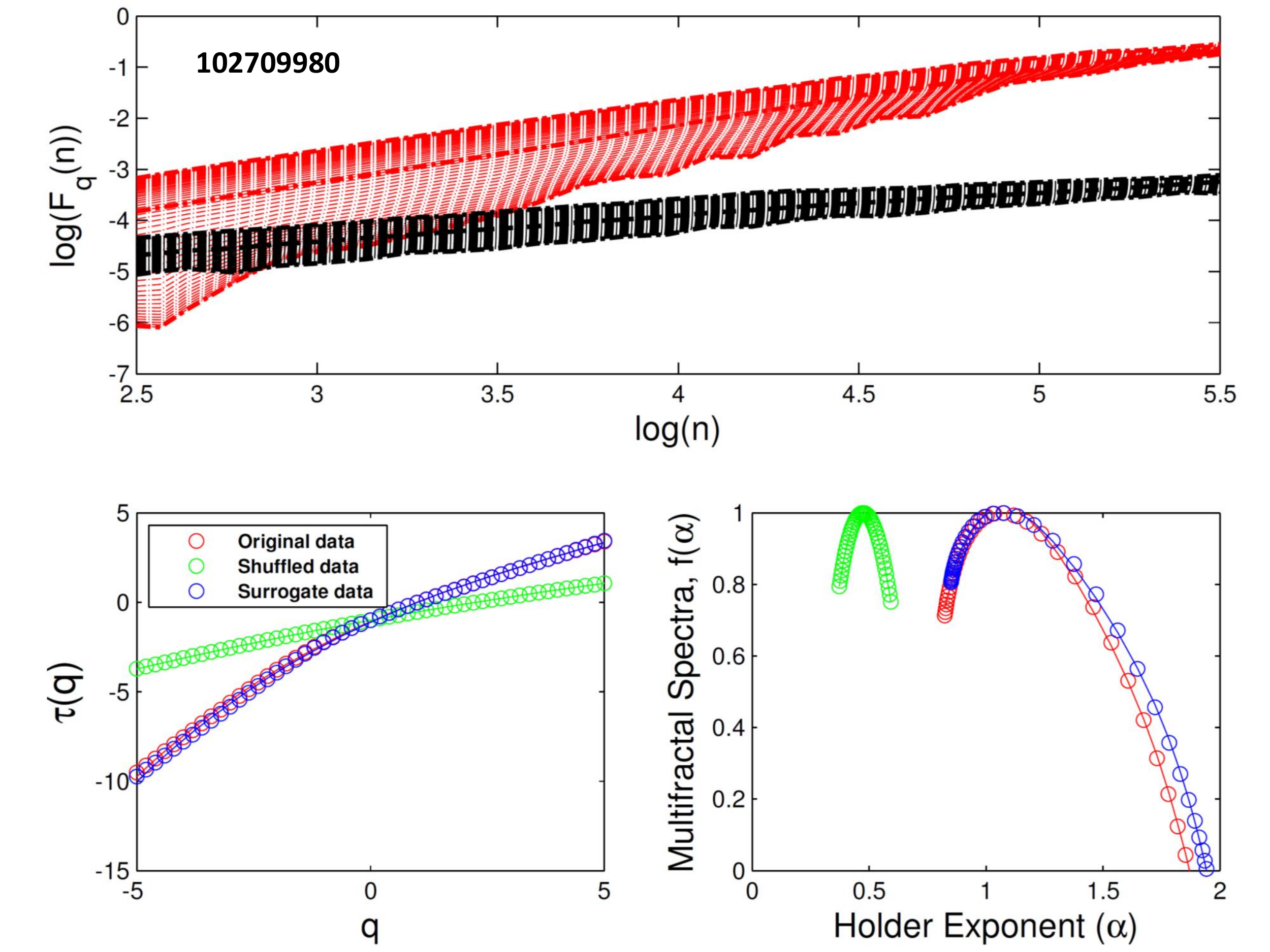}}
\subfigure{\includegraphics[width=0.48\textwidth]{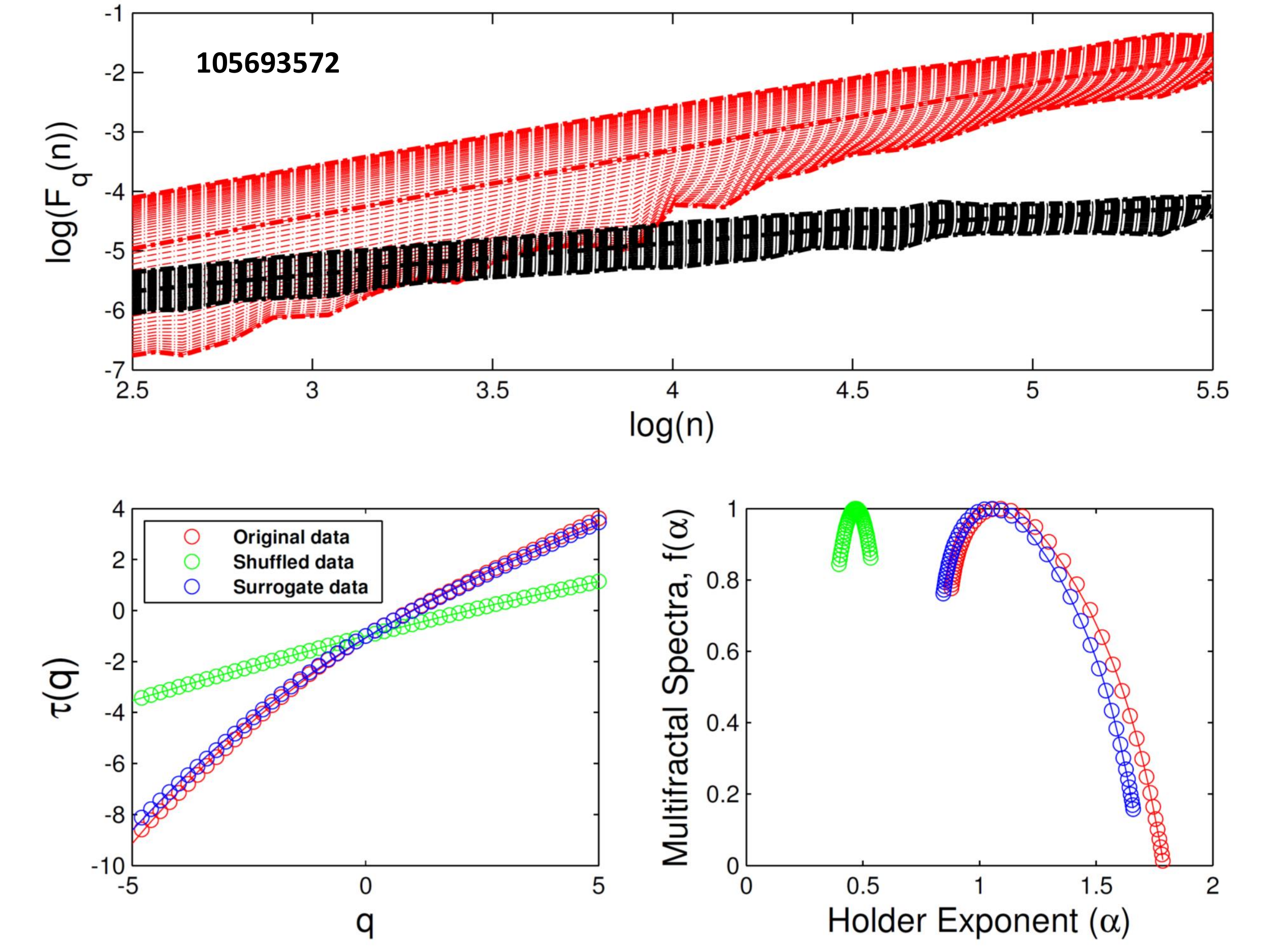}}
\end{center}
\caption{Same procedure proposed in figure \ref{fig1} for the New Suns sample: CoRoT ID 102709980 (left panel) and 105693572 (right panel). As cited by \cite{defreitas2013}, CoRoTID 105693572 is the best candidate for New Sun. 
}
\label{fig2}
\end{figure*}

\begin{figure*}
\begin{center}
\subfigure{\includegraphics[width=0.48\textwidth]{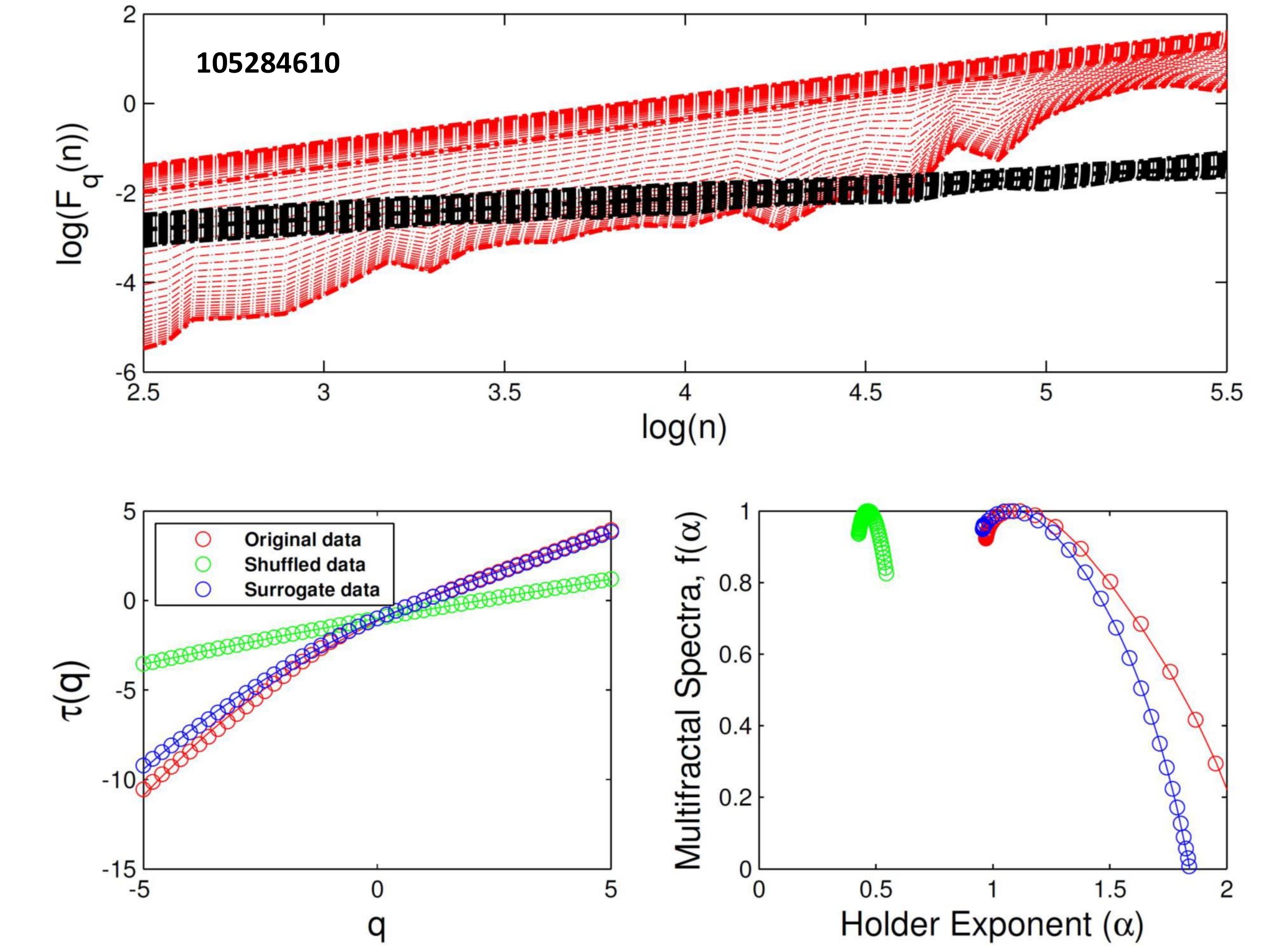}}
\subfigure{\includegraphics[width=0.48\textwidth]{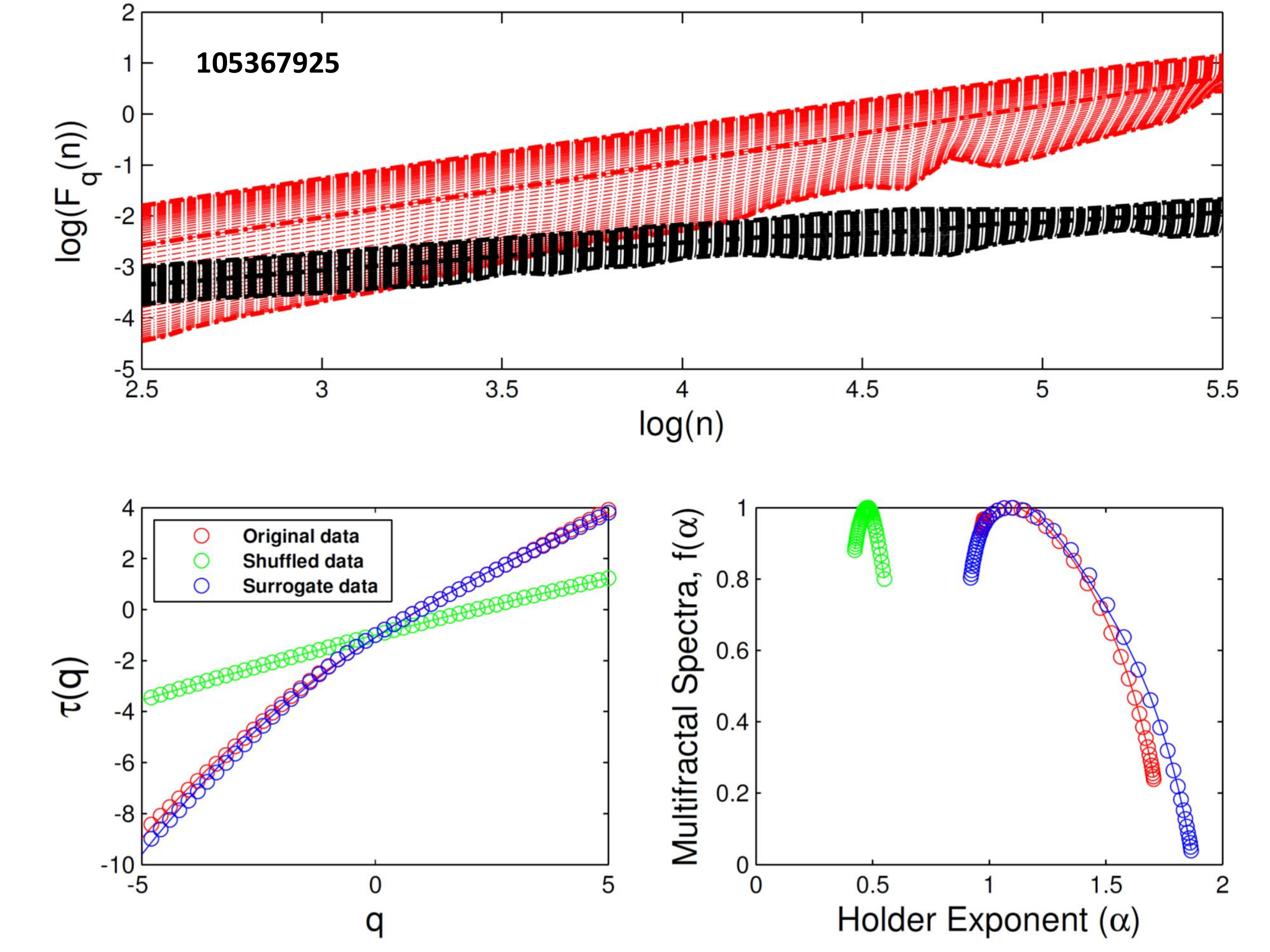}}
\end{center}
\caption{Same procedure proposed in figure \ref{fig1} for the super-solar sample named by CoRoT ID 105284610 (left panel) and 105367925 (right panel). 
}
\label{fig3}
\end{figure*}

\begin{figure*}
\begin{center}
\subfigure{\includegraphics[width=0.48\textwidth]{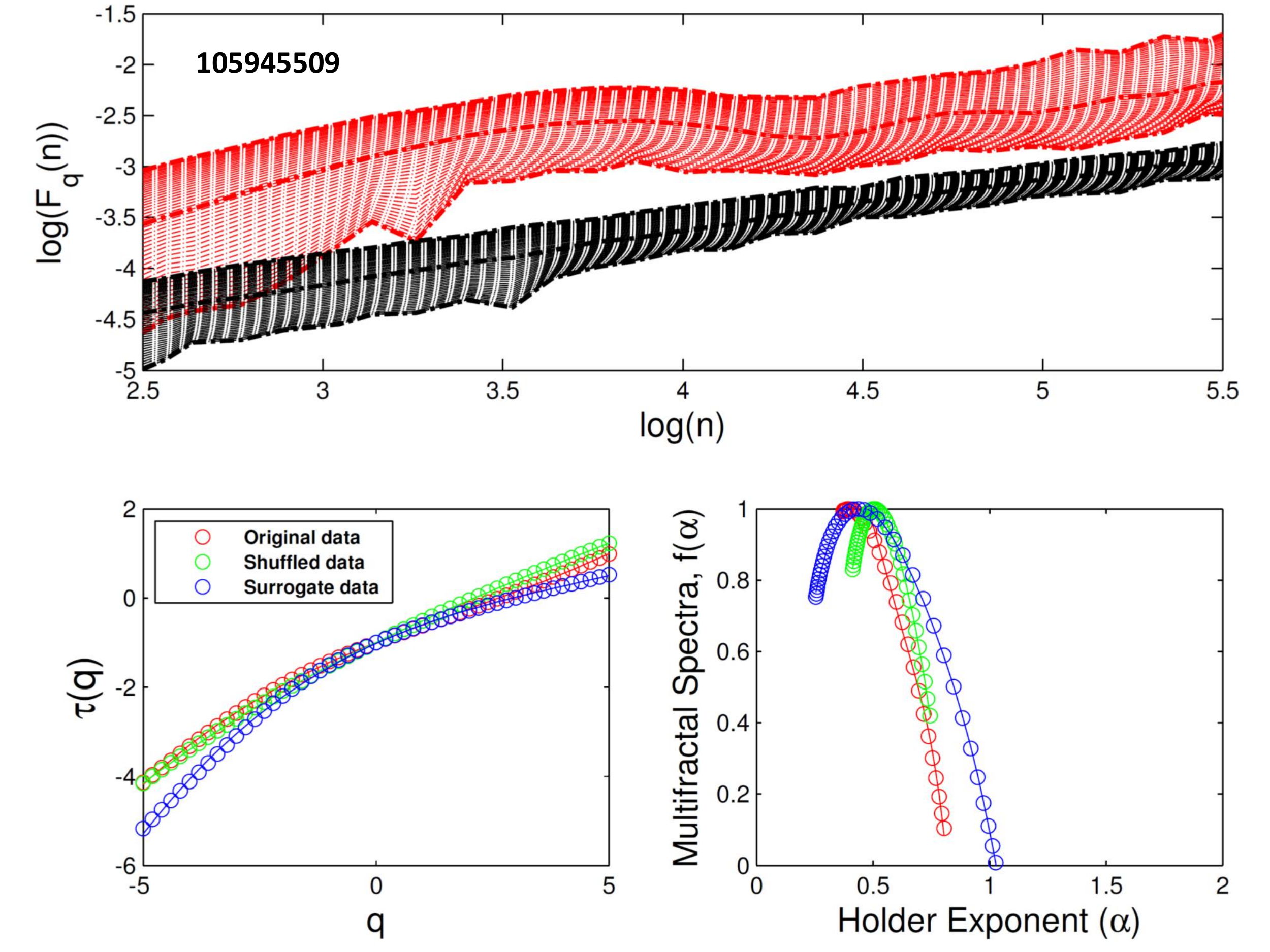}}
\subfigure{\includegraphics[width=0.48\textwidth]{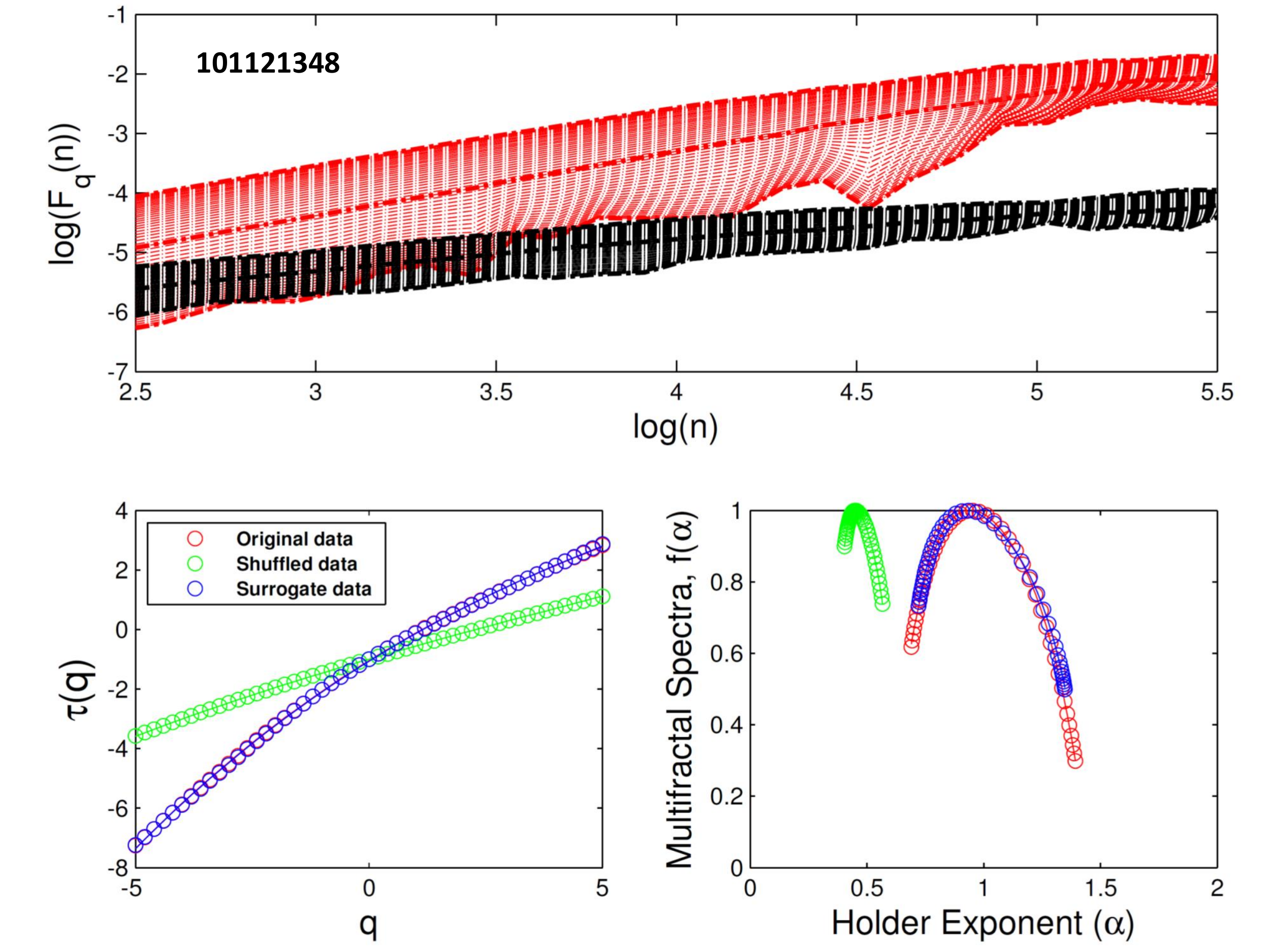}}
\end{center}
\caption{Same procedure proposed in figure \ref{fig1} for the sub-solar sample named by CoRoT ID 105945509 (left panel) and 101121348 (right panel). 
}
\label{fig4}
\end{figure*}

\begin{figure*}
\begin{center}
\subfigure{\includegraphics[width=0.48\textwidth]{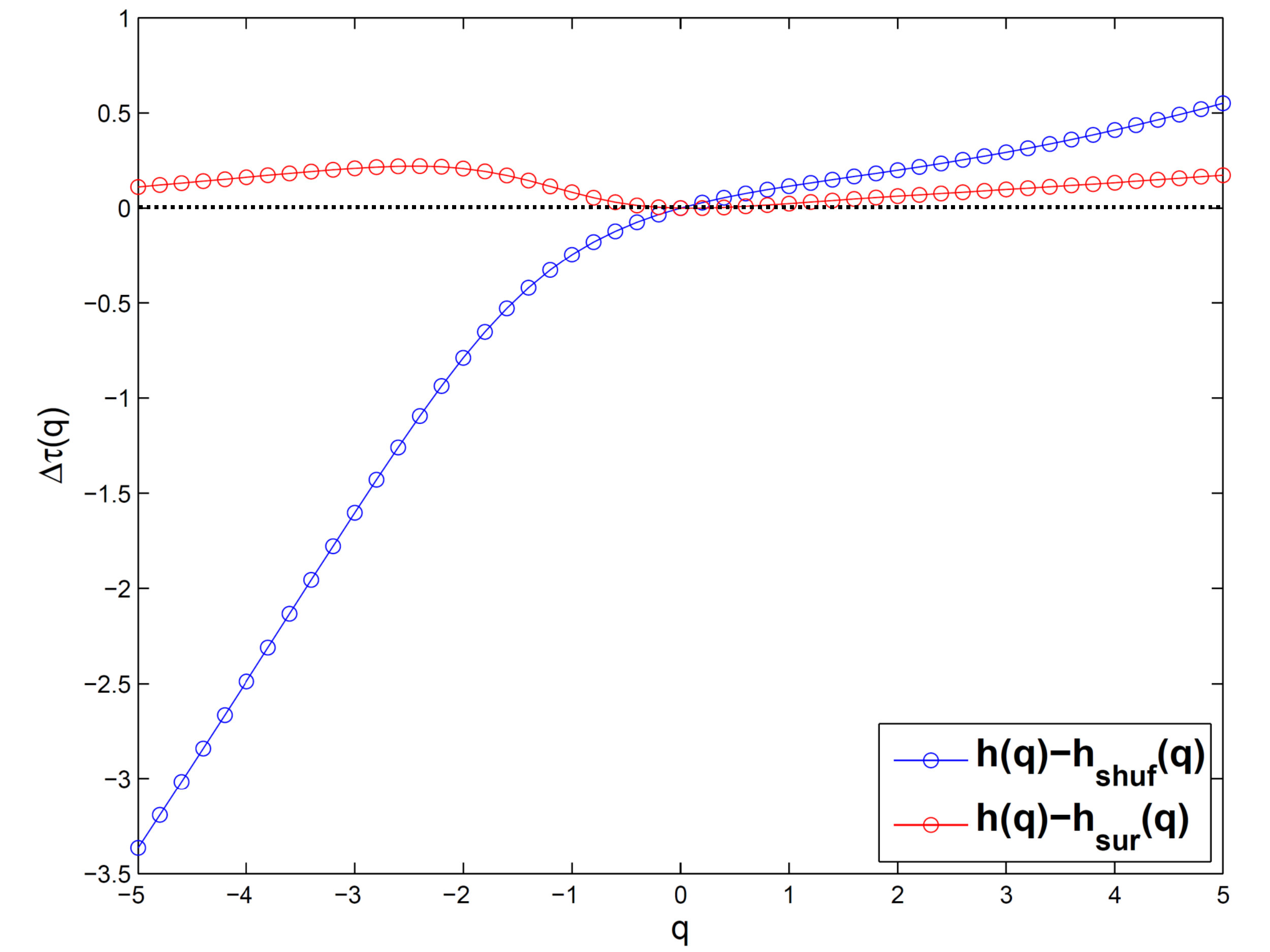}}
\subfigure{\includegraphics[width=0.48\textwidth]{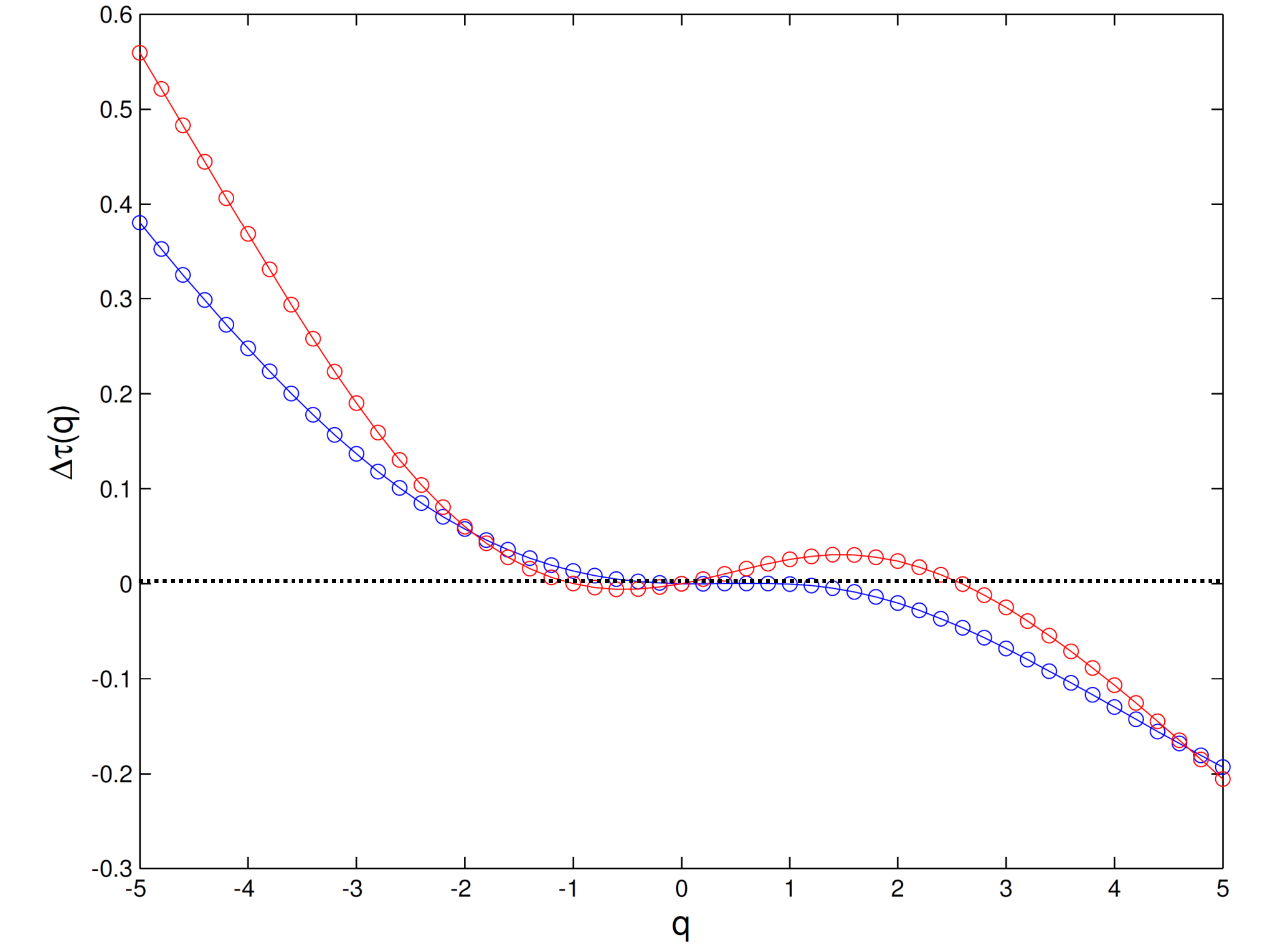}}
\end{center}
\caption{Deviations $\Delta\tau(q)$ as a function of $q$ between the fluctuation functions with the active (left panel) and quiet (right panel) Sun original data, respectively, due to long-range correlations and broadness of probability distribution. 
}
\label{fig6}
\end{figure*}

\begin{figure*}
\begin{center}
\subfigure{\includegraphics[width=0.48\textwidth]{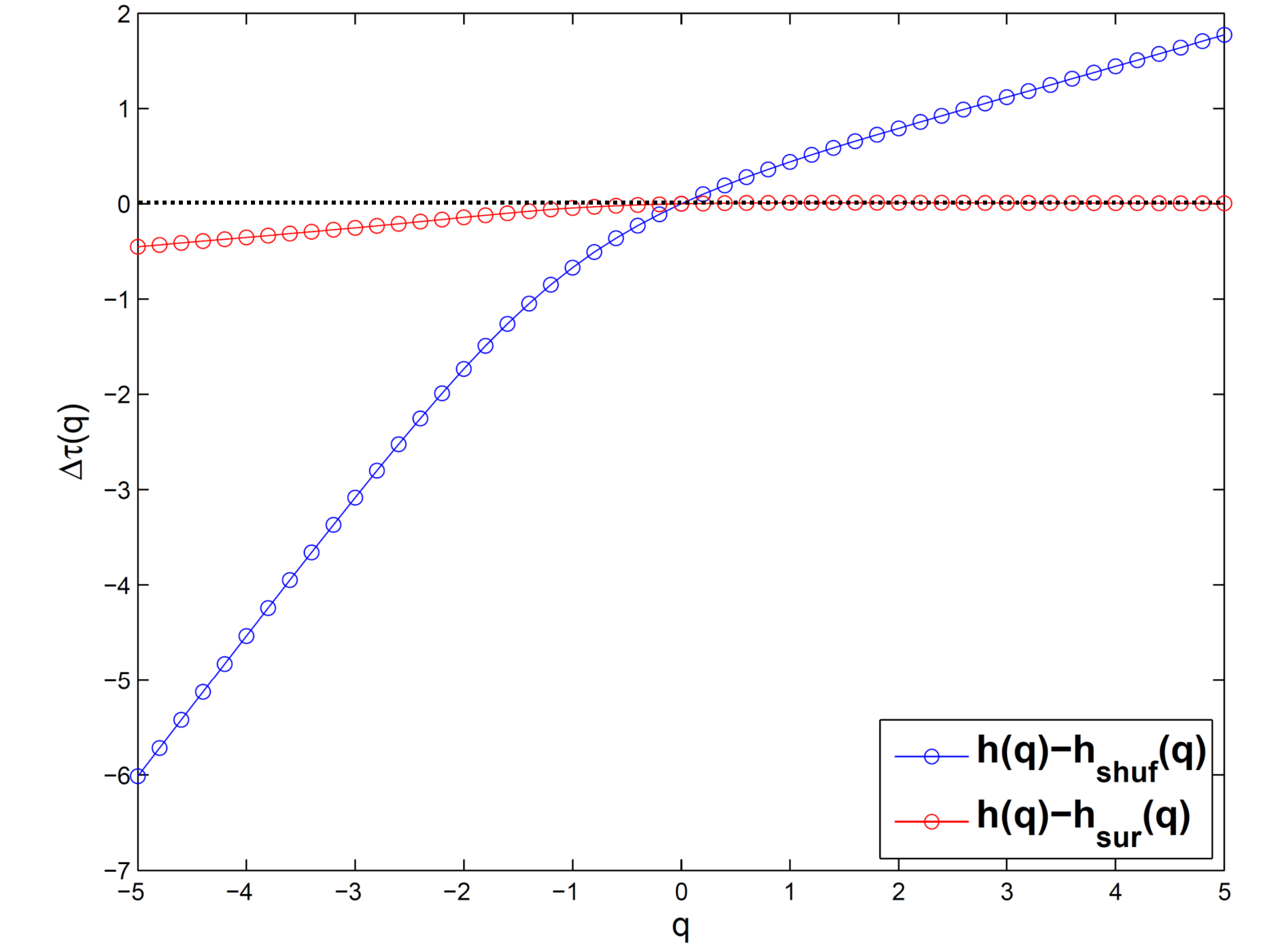}}
\subfigure{\includegraphics[width=0.48\textwidth]{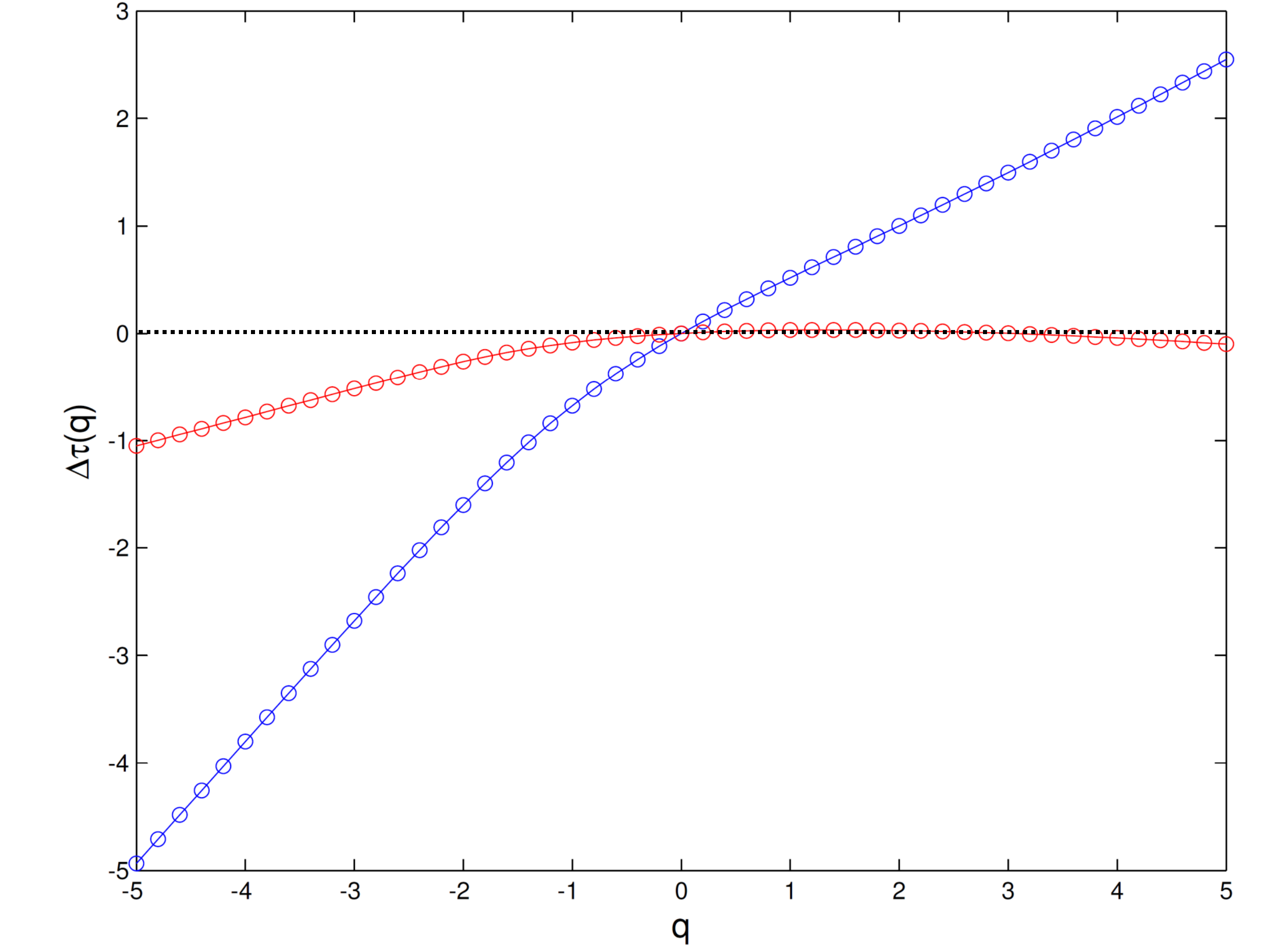}}\\
\subfigure{\includegraphics[width=0.48\textwidth]{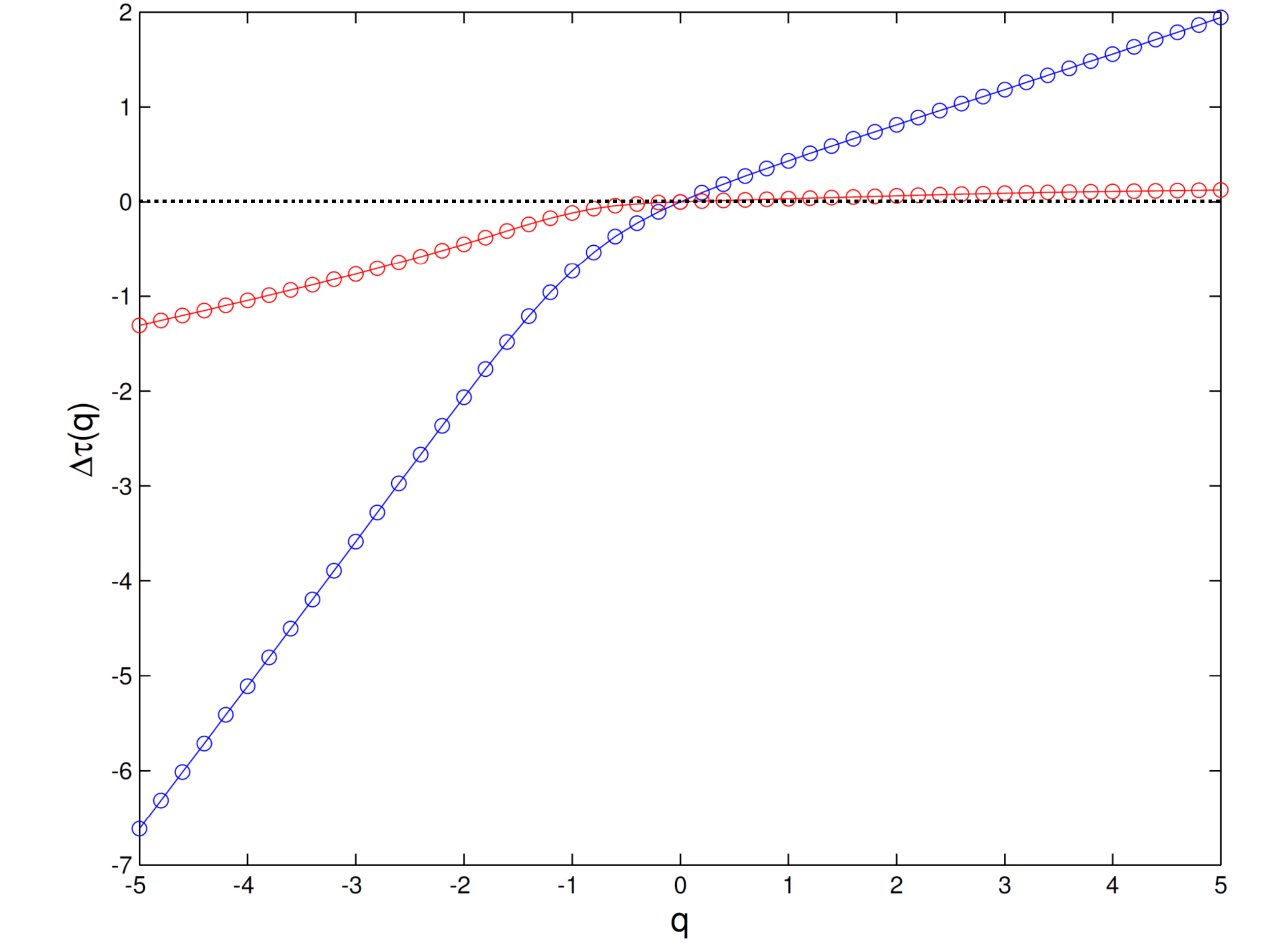}}
\subfigure{\includegraphics[width=0.48\textwidth]{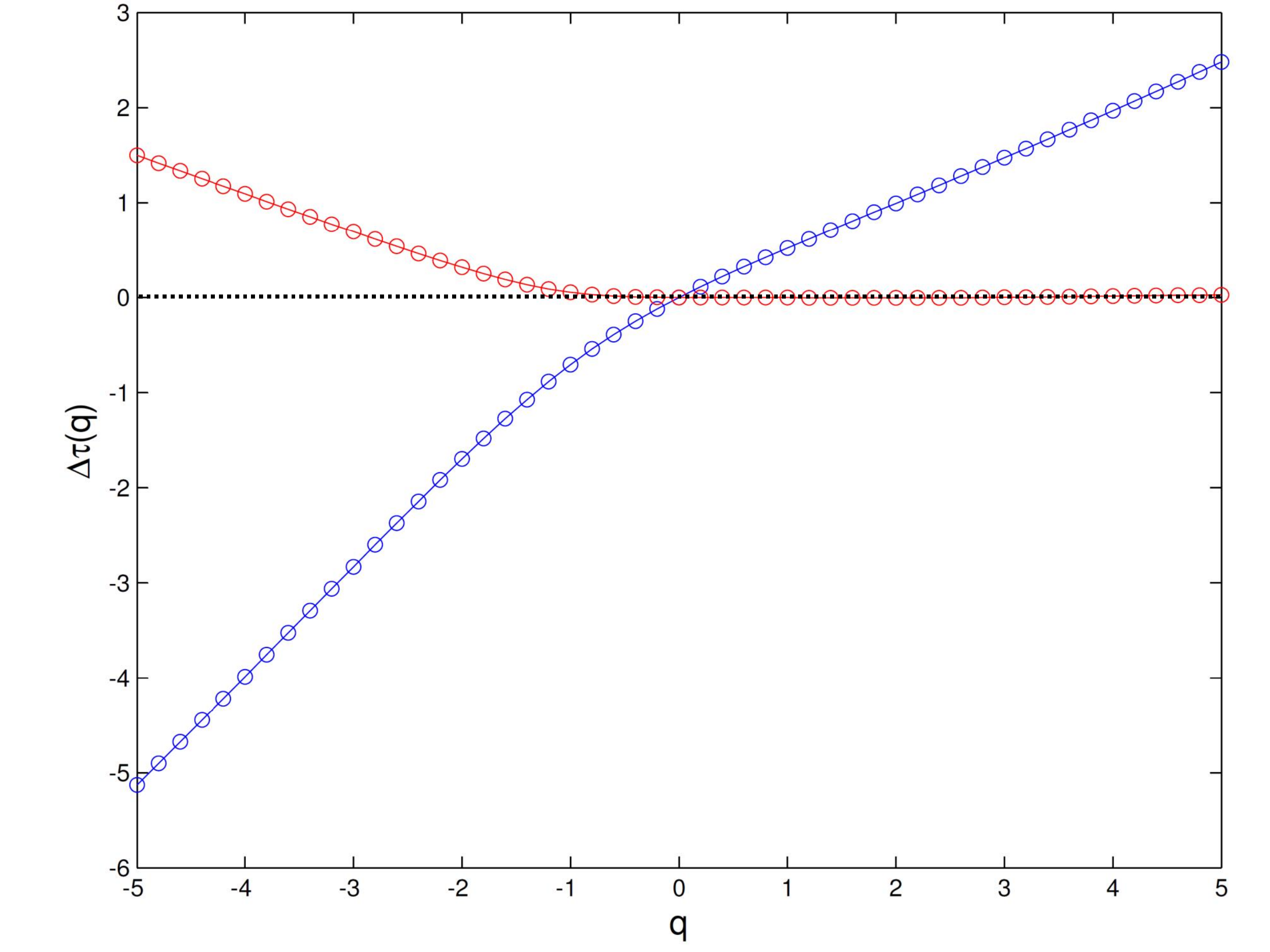}}\\
\subfigure{\includegraphics[width=0.48\textwidth]{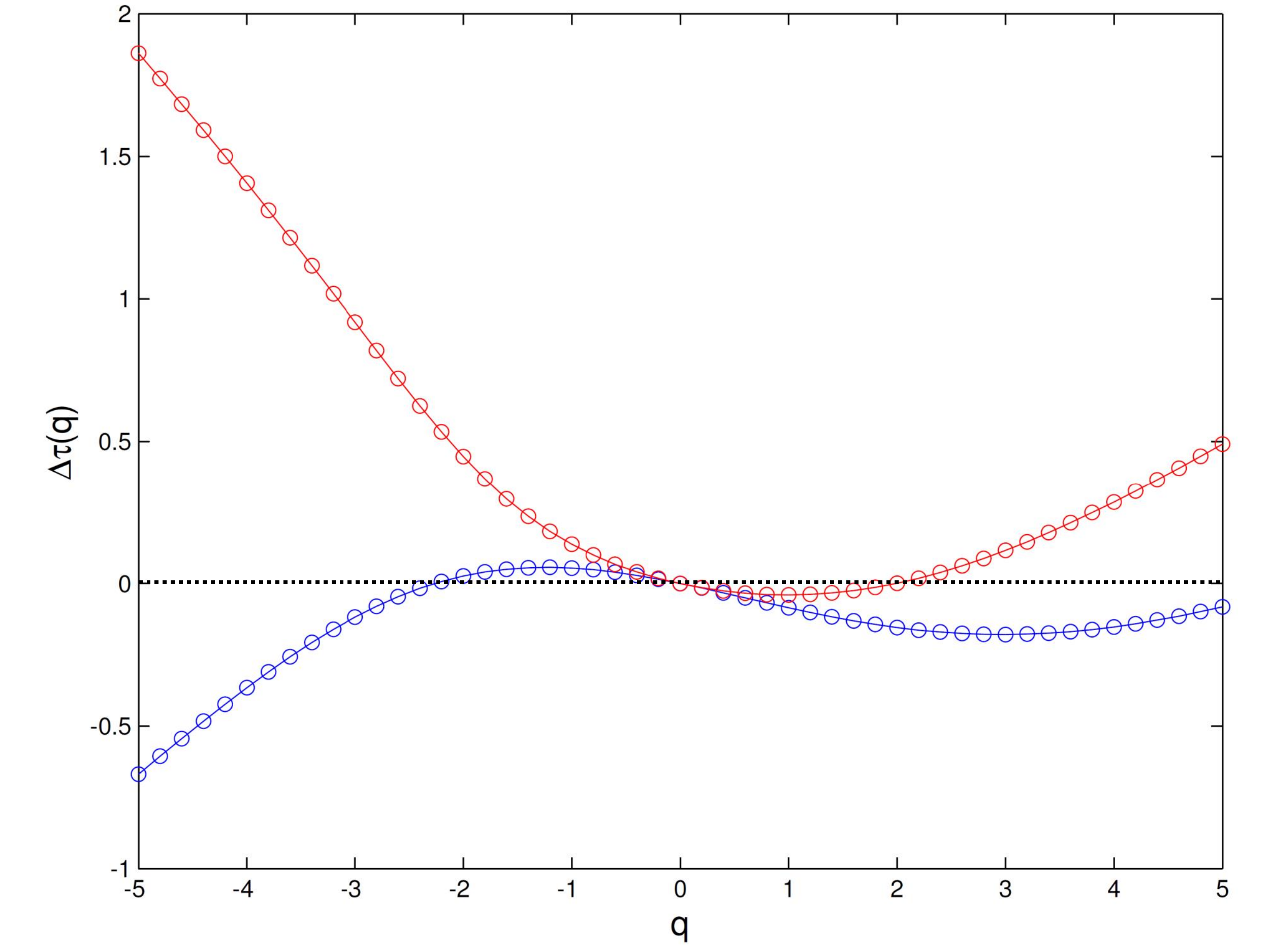}}
\subfigure{\includegraphics[width=0.48\textwidth]{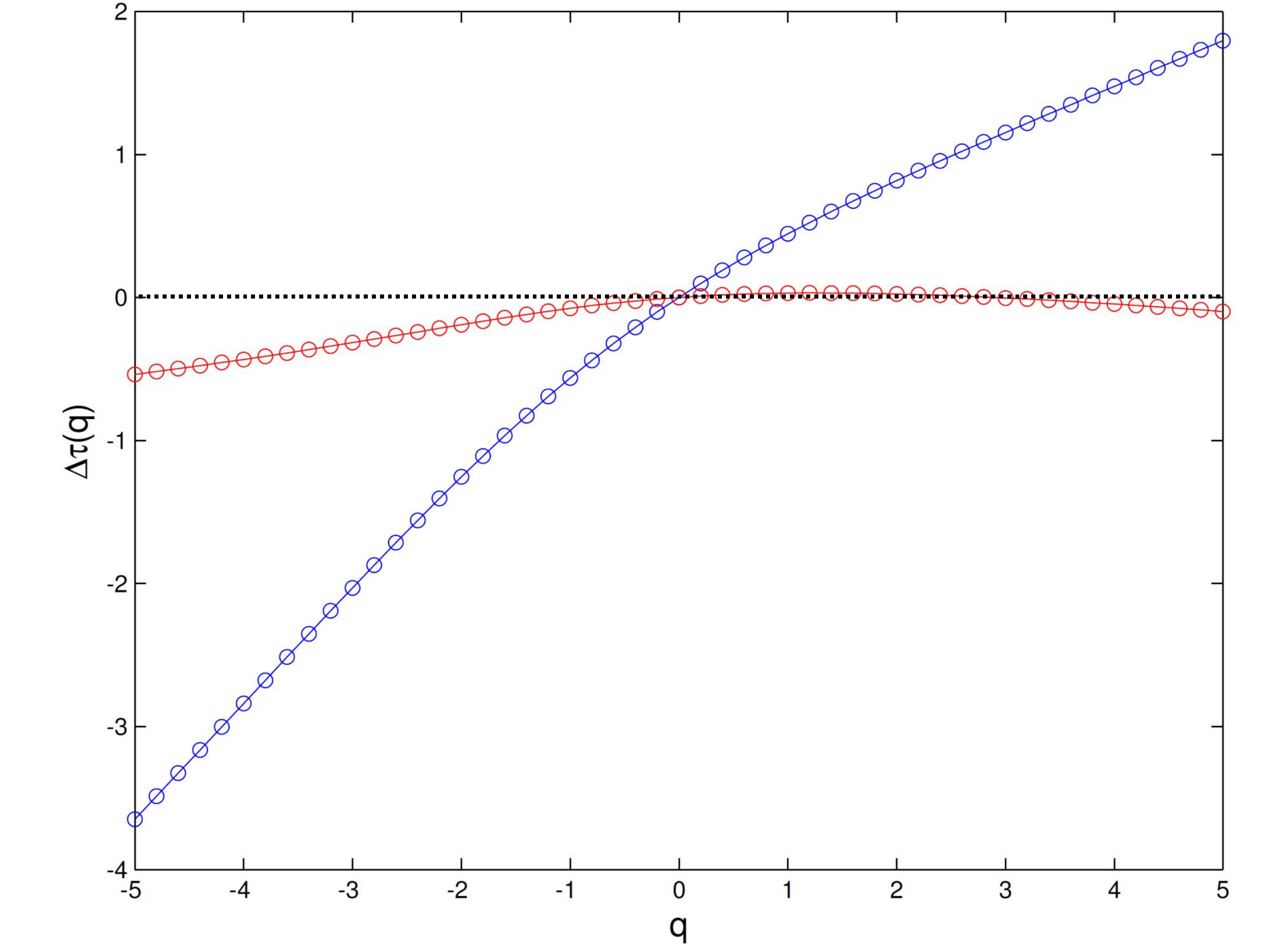}}
\end{center}
\caption{Same procedure proposed in figure \ref{fig6} according to the sequence of the named stars in Figures \ref{fig2}, \ref{fig3} and \ref{fig4}.
}
\label{fig7}
\end{figure*}

As shown in Figs. \ref{fig1} through \ref{fig4}, the long left tails are related to the dominant noise amplitude, whereas the long right tails are associated with the dominance of the rotational modulation amplitude. Furthermore, these results indicate that the variability due to rotational modulation is even more complex and dynamic than is indicated by classical statistical methods, such as Fourier analysis and the autocorrelation function.  

\begin{figure*}
\begin{center}
\subfigure{\includegraphics[width=0.7\textwidth]{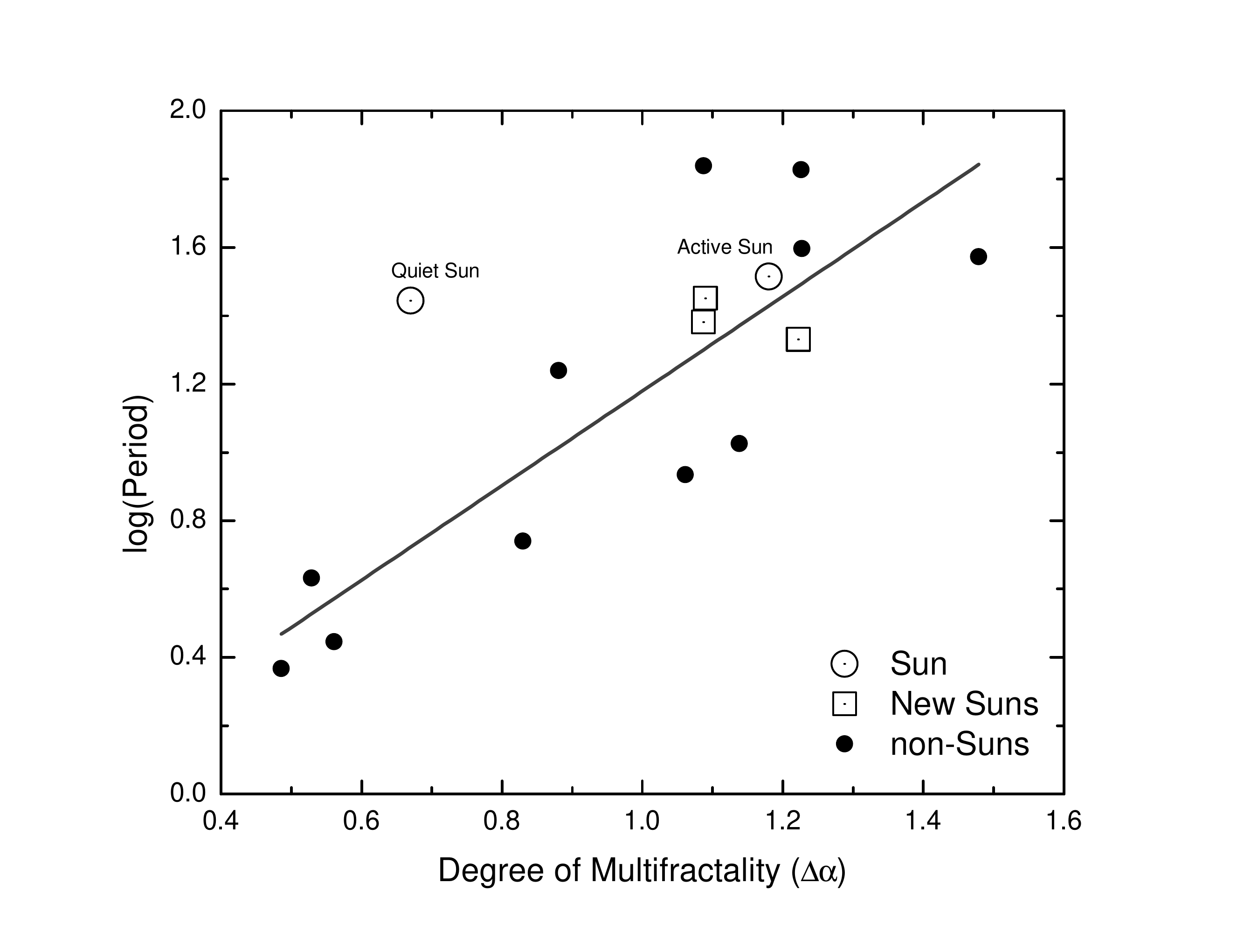}}\\
\subfigure{\includegraphics[width=0.7\textwidth]{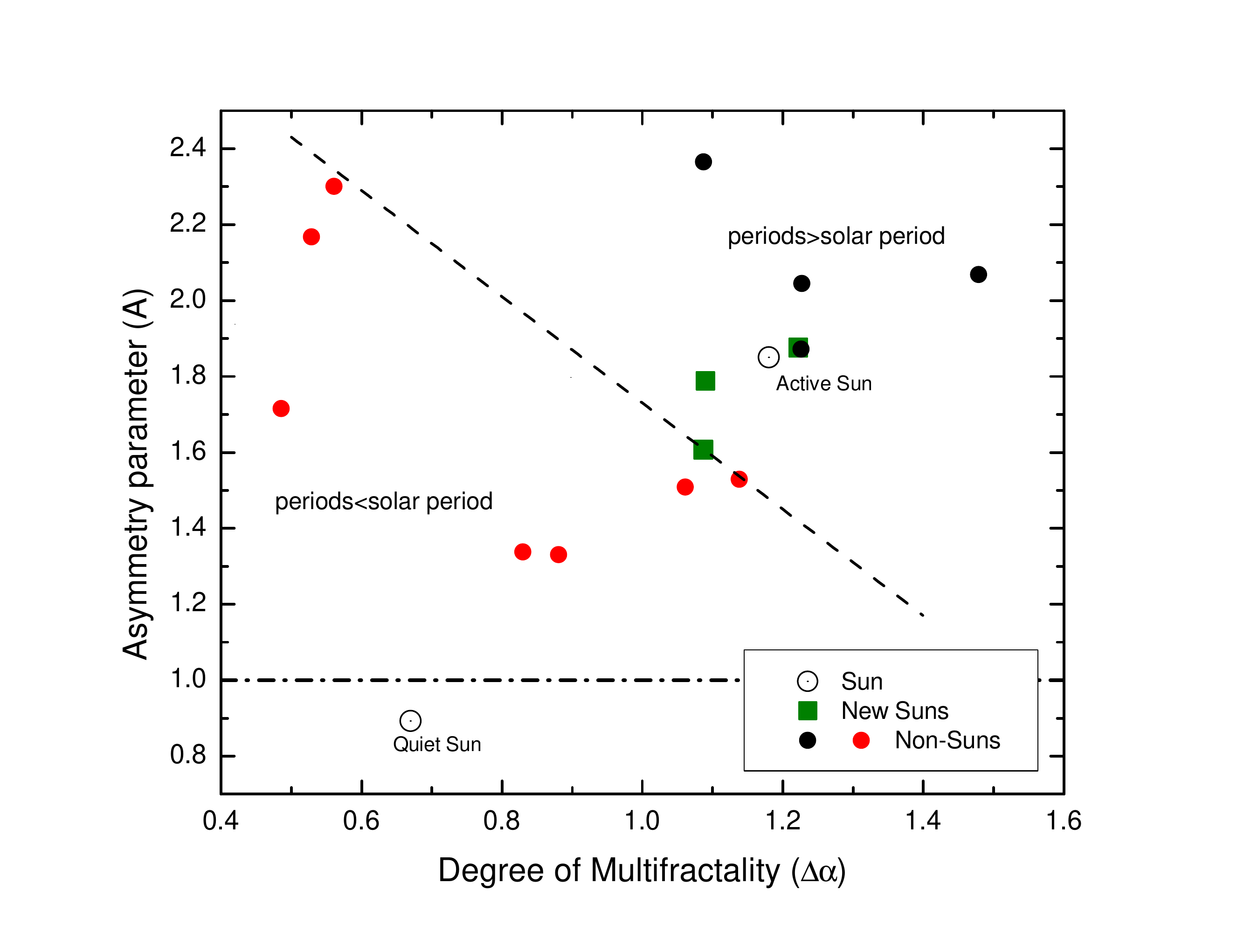}}
\end{center}
\caption{Plotted values of the degree of multifractality ($\Delta\alpha$), derived on the basis of multifractal analysis, as a function of rotation period. The Sun was analyzed in its active and quiet phases. The solid line denotes the linear regression with the very similar adjustment to Equation 1 derived by \cite{defreitas2013} ($top$). $\Delta\alpha$ plotted as a function of the asymmetry parameter $A$. The sloping line indicates the separation in two rotational regimes, while the horizontal line separate the stars in two multifractal asymmetry regimes (bottom). 
}
\label{fig5}
\end{figure*}

\begin{figure*}
\begin{center}
\subfigure{\includegraphics[width=0.7\textwidth]{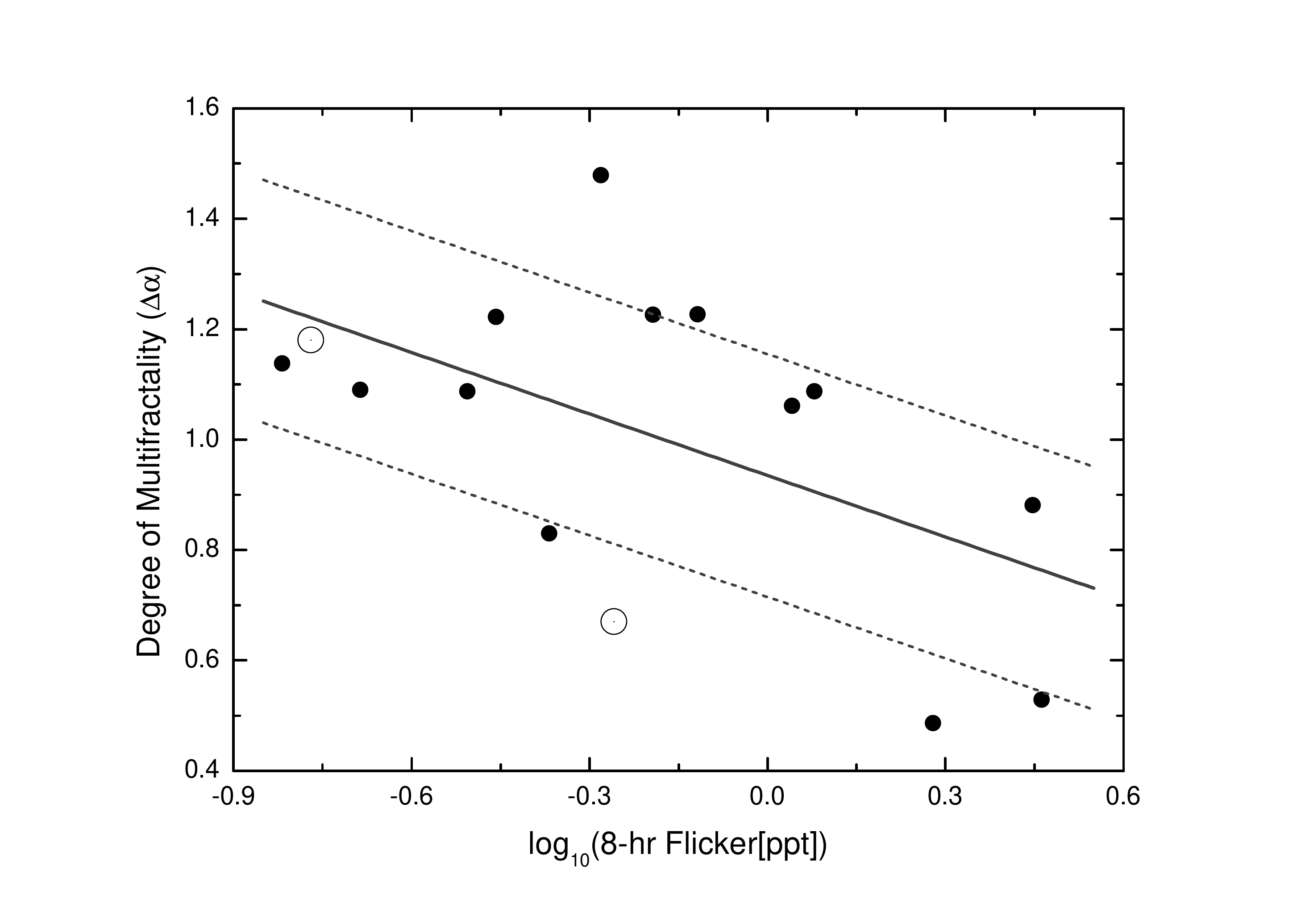}}
\end{center}
\caption{Relationship between the degree of multifractality ($\Delta\alpha$) and the 8-hr flicker noise. Solid line represents the best-fitting linear regression, while short dashed lines show the 95 percent confidence level.
}
\label{fig8}
\end{figure*}

The analysis of the behavior of the rotation period as a function of the degree of multifractality suggests that the rotation period of stars is inherently scaled by a dynamical change from homogeneity toward heterogeneity, reciprocally, from shorter to larger periods, as emphasized by the growth in multifractality shown in Fig. \ref{fig5} (see the top panel). This behavior can be related to strong magnetic activity, thus indicating that the multifractal degree is a suitable activity indicator. The method used by \cite{defreitas2013} derived a single relation that was in good agreement with the results presented here. Nevertheless, our results highlight that the multifractal analysis has led to a better understanding of how such deterministic and stochastic characteristics are regulated hierarchically.

Another result, but one that requires further studies, is related to the asymmetry of the multifractal spectrum. If this asymmetry is confirmed, the transition region between left and right asymmetries could be used to discriminate between stars in the active and quiet phases. If this hypothesis is true, our work could lead us to assume that each CoRoT star is in the active phase. However, more realistic models and methods must be developed to understand the behavior of astrophysical noise and the fluctuations due to the starspot lifetime. Finally, our multifractal approach can be a useful tool for understanding the dynamical mechanisms that control stellar magnetic activity.

\acknowledgments
Research activities of the Astronomy Observational Board of the Federal University of
Rio Grande do Norte are supported by continuous grants from CNPq and 
FAPERN brazilian agencies. We also acknowledge 
financial support from INCT INEspa\c{c}o/CNPq/MCT. MLC, JPB and ADC acknowledge a CAPES/PNPD fellowship. MMFN and PRVM acknowledge graduate fellowships from 
CAPES. DBF  also acknowledges financial support by the Brazilian agency CNPq (Grant No. 306007/2015-0).

\begin{deluxetable}{lccccccc}
\tabletypesize{\scriptsize}
\tablecaption{Rotation period of our stellar sample and the results of the multifractal analysis.}
\tablewidth{0pt}
\tablehead{{\bf Star} & Broadness ($\Delta\alpha$) & Skewness ($A$) & $\alpha_{0}$ &  $\log$Per & $\delta Per$\\
 &  &  &  & (days) &(days) 
}
\startdata
\multicolumn{6}{c}{New Sun candidates as Compared with the Sun}\\
\hline
{\bf Active Sun}	&	1.18	&	 1.85 &	1.07	&	1.52  & 0.4	\\ 
{\bf Quiet Sun}	&	0.67	&	0.89	&	1.02	& 1.44  & 0.3 \\ 
{\bf ID  100746852 }	&	1.09	&	1.61	&	1.08	&	1.38 & 0.3 \\ 
{\bf ID  102709980 }	&	1.22	&	1.88	&	1.07	&	1.33 & 0.1\\ 
{\bf ID  105693572 }	&	1.09	&	1.79	&	1.09	&	1.45 & 0.5\\ 
\hline
\multicolumn{6}{c}{Comparison Stars}\\
\hline
\multicolumn{6}{c}{Super-Solar Sample}\\
\hline
{\bf ID 105085209}	&	1.23	&	2.04	&	1.12 &		1.60 & 0.9\\
{\bf ID 105284610}	&	1.48	&	2.07	&	1.12&		1.57 & 0.4	\\
{\bf ID 105367925}	&	1.09	&	2.37	&	1.01	&		1.84 & 1.0	\\
{\bf ID 105379106}	&	1.23	&	1.87	&	1.09	&		1.83 & 1.0	\\
\hline
\multicolumn{6}{c}{Sub-Solar Sample}\\
\hline
{\bf ID 101121348}	&	0.88	&	1.33	&	0.95	&		1.24& 0.2	\\
{\bf ID 101710670}	&	0.83	&	1.33	&	0.90	&		0.74& 0.02	\\
{\bf ID 102752622}	&	0.49	&	1.72	&	0.31	&		0.37 & 0.0003	\\
{\bf ID 102770893}	&	0.53	&	2.17	&	0.73	&		0.63 & 0.002	\\
{\bf ID 105503339}	&	1.14	&	1.53	&	0.99	&		1.03 & 0.4	\\
{\bf ID 105945509}	&	0.56	&	2.30  &	0.43	&		0.46 & 0.003	\\
{\bf ID 105845539 }	&	1.06	&	1.51	&	1.02	&		0.94	& 0.4 \\ 
 \enddata
\label{tab1}
\end{deluxetable}

\begin{thebibliography}{}

\bibitem[\protect\citeauthoryear{Alessio et al.}{2002}]{alessio}
Alessio, E, Carbone, A., Castelli, G., \& Frappietro, V. 2002, Eur. Physi. J., 27, 197

\bibitem[\protect\citeauthoryear{Aschwanden \& Parnell}{2002}]{aschwa}
Aschwanden, M. J., \& Parnell, C. E. 2002, Astrophys. J., 572, 1048

\bibitem[\protect\citeauthoryear{Aschwaden}{2011}]{a2011}
Aschwanden, M. J. 2011, Self-Organized Criticality in Astrophysics. The Statistics
of Nonlinear Processes in the Universe, Springer-Praxis: New York

\bibitem[\protect\citeauthoryear{Baglin}{2006}]{baglin2006}
Baglin, A. 2006, in ESA Special Publication, Vol. 1306, ed. M. Fridlund, A. Baglin, J. Lochard, \& L. Conroy, 111

\bibitem[\protect\citeauthoryear{Bastien et al.}{2013}]{bastien}
Bastien, F. A., Stassun, K. G., Basri, G., \& Pepper, J. 2013, Nature, 500, 427
 
\bibitem[\protect\citeauthoryear{Bravo et al.}{2011}]{bravo}
Bravo, J. P., Roque, S., Estrela, R., Le\~ao, I. C., \& De Medeiros, J. R. 2014, A\&A, 568, 34

\bibitem[\protect\citeauthoryear{Carter \& Winn}{2009}]{cw2009}
Carter, J. A., \& Winn, J. N. 2009, \apj, 704, 51

\bibitem[\protect\citeauthoryear{Chappell \& Scalo}{2001}]{cs2001}
Chappell, D., \& Scalo, J., 2001, \apj, 551, 712.

\bibitem[\protect\citeauthoryear{De Medeiros et al.}{2013}]{demedeiros2013}
De Medeiros, J. R., Lopes, C. E. F., Le\~ao, I. C., et al. 2013, \aap, 555, 63

\bibitem[\protect\citeauthoryear{de Freitas \& De Medeiros}{2009}]{defreitas2009}
de Freitas, D. B., \& De Medeiros, J. R. 2009, Europhys. Lett, 88, 19001

\bibitem[\protect\citeauthoryear{de Freitas et al.}{2013a}]{defreitas2013b}
de Freitas, D. B., Fran\c{c}a, G. S., Scherrer, T. M., Vilar, C. S., \& Silva, R. 2013a, Europhys. Lett., 102, 39001

\bibitem[\protect\citeauthoryear{de Freitas et al.}{2013b}]{defreitas2013}
de Freitas, D. B., Le\~ao, I. C., J. R., Lopes, C. E. F., De Medeiros, J. R., et al. 2013b, ApJL, 773, L18

\bibitem[\protect\citeauthoryear{Feder}{1988}]{feder1988}
Feder, J. 1988, Fractals, Plenum Press, New York    

\bibitem[\protect\citeauthoryear{Feigelson \& Babu}{2012}]{feigelson}
Feigelson, E. D., \& Jogesh Babu, G. 2012, Modern Statistical Methods for Astronomy,
Cambridge University Press (Cambridge, UK)

\bibitem[\protect\citeauthoryear{Gu \& Zhou}{2010}]{gu2010}
Gu, G.-F., \& Zhou, W.-X. 2010, Phys. Rev. E, 82, 011136

\bibitem[\protect\citeauthoryear{Grech \& Pamula}{1982}]{gp}
Grech, D., \& Pamula, G. 2008, Physica A, 387, 4299


\bibitem[\protect\citeauthoryear{Hilhorst}{2009}]{Hilhorst}
Hilhorst, H. J. 2009, Brazilian Journal of Physics, 39, 2

\bibitem[\protect\citeauthoryear{Hurst}{1951}]{hurst1951}
Hurst, H. E. 1951, Trans. Am. Soc. Civ. Eng., 116, 770

\bibitem[\protect\citeauthoryear{Hurst, Black \& Simaika}{1965}]{hurst1965}
Hurst, H. E. \& Black, R. P., \& Simaika, Y. M. 1965, Long-term storage: an experimental study, Constable, London

\bibitem[\protect\citeauthoryear{Ihlen}{2012}]{ihlen}
Ihlen, E. A. F. 2012, Front. Physiology 3, 141

\bibitem[\protect\citeauthoryear{Ivanov et al.}{1999}]{ivanov1999}
Ivanov, P. Ch., Amaral, L. A. N., Goldberger, A. L., et al. 1999, \nat, 399, 461

\bibitem[\protect\citeauthoryear{Kipping et al.}{2014}]{kipping}
Kipping, D. M., Bastien, F. A., Stassun, K. G., Chaplin, W. J., Huber, D. \& Buchhave, L. A. 2014, ApJL, 785, 32

\bibitem[\protect\citeauthoryear{Kraft}{1967}]{kraft1967}
Kraft, R. P. 1967, ApJ, 150, 551

\bibitem[\protect\citeauthoryear{Kantelhardt et al.}{2002}]{Kantelhardt}
Kantelhardt, J.W., Zschiegner, S.A., Koscienlny-Bunde, E., Havlin, S., Bunde, S. A., \& Stanley,  H. E. 2002, Physica A, 316, 87

\bibitem[\protect\citeauthoryear{Karoff et al.}{2013}]{karoff}
Karoff, C., Campante, T. L., Ballot, J., Kallinger, T., Gruberbauer, M., et al. 2013, ApJ, 767, 34

\bibitem[\protect\citeauthoryear{Komm}{1995}]{komm1995}
Komm, R. W. 1965, Solar Phys., 156, 17

\bibitem[\protect\citeauthoryear{Lanza et al.}{2003}]{lanza2003}
Lanza, A. F., Rodono, M., Pagano, I., et al. 2003, \aap, 403, 1135

\bibitem[\protect\citeauthoryear{Lanza, Rodono \& Pagano}{2004}]{lanza2004}
Lanza, A. F., Rodono, M., Pagano, I. 2003, A\&A, 425, 707


\bibitem[\protect\citeauthoryear{Lomb}{1976}]{lomb1976}
Lomb, N. R. 1976, \apss, 39, 447

\bibitem[\protect\citeauthoryear{Del Moro}{2004}]{moro}
Del Moro, D. 2004, A\&A, 428, 1007

\bibitem[\protect\citeauthoryear{Mathur et al.}{2014}]{mathur}
Mathur, S., Garcia, R. A., Ballot, J., Ceillier, T., Salabert, D., et al. 2014, A\&A, 562, 124

\bibitem[\protect\citeauthoryear{Movahed et al.}{2006}]{movahed}
Movahed, M.S., Jafari, G.R., Ghasemi, F., Rahvar, S., \& Reza, M. R. T. 2006. Stat. Mech., 02003

\bibitem[\protect\citeauthoryear{Mandelbrot \& Wallis}{1969a}]{mw1969a}
Mandelbrot, B., \& Wallis, J. R. 1969a, Water Resour. Res., 5, 521

\bibitem[\protect\citeauthoryear{Mandelbrot \& Wallis}{1969b}]{mw1969b}
Mandelbrot, B., \& Wallis, J. R. 1969b, Water Resour. Res., 5, 967

\bibitem[\protect\citeauthoryear{Mandelbrot \& Wallis}{1969c}]{mw1969c}
Mandelbrot, B., \& Wallis, J. R. 1969b, Water Resour. Res., 5, 967

\bibitem[\protect\citeauthoryear{Muzy et al.}{1991}]{muzy1991}
Muzy, J.-F., Bacry, E., \& Arneodo,  A. 1994, Phys. Rev. Lett. 67, 3515

\bibitem[\protect\citeauthoryear{Muzy et al.}{1994}]{muzy1994}
Muzy, J.-F., Bacry, E., \& Arneodo,  A. 1994, Int. J. Bifurc. Chaos, 4, 245

\bibitem[\protect\citeauthoryear{Norouzzadeha, Dullaertc \& Rahmani} {2007}]{Norouzzadeha}
Norouzzadeha, P., Dullaertc, W., \& Rahmani, B. 2007, Physica A, 380, 333

\bibitem[\protect\citeauthoryear{Press et al.}{2007}]{press}
Press, W. H., Saul A. T., William T. V., \& Brian P. F. 2007, Numerical Recipes in C: The Art of Scientific Computing, Third Edition, Cambridge University Press.

\bibitem[\protect\citeauthoryear{Radick et al.}{1998}]{radick}
Radick, R. R., Lockwood, G. W., Skiff, B. A., \& Baliunas, S. L. 1998, ApJSS, 118, 239

\bibitem[\protect\citeauthoryear{Ruan \& Zhou}{2011}]{ruan}
Ruan, Y.-P, \& Zhou, W.-X. 2011, Physica A, 390, 3512

\bibitem[\protect\citeauthoryear{Sarro et al.}{2013}]{sarro2013}
Sarro, L. M., Debosscher, J., Neiner, C., et al. 2013, \aap, 550, 120

\bibitem[\protect\citeauthoryear{Scargle}{1982}]{scargle1982}
Scargle, J. D. 1982, \apj, 263, 835

\bibitem[\protect\citeauthoryear{Sen}{2007}]{sen2007}
Sen, A. K. 2007, Solar Phys, 241, 67

\bibitem[\protect\citeauthoryear{Suyal, Prasad \& Singh}{2009}]{sps2009}
Suyal, V., Prasad, A., \& Singh, H. P. 2009, Solar Phys., 260, 441

\bibitem[\protect\citeauthoryear{Pascual-Granado}{2011}]{pg2011}
Pascual-Granado, J. 2011, Highlights of Spanish Astrophysics VI, Proceedings of the IX Scientific Meeting of the Spanish Astronomical Society (SEA), held in Madrid, September 13 - 17, 2010, Eds.: M. R. Zapatero Osorio, J. Gorgas, J. Maíz Apellániz, J. R. Pardo, and A. Gil de Paz., p. 744

\bibitem[\protect\citeauthoryear{Tang et al.}{2015}]{tang}
Tang, L., Lv, H., Yang, F., \& Yu, L. 2015, Chaos, Solitons \& Fractals, 81, 117

\bibitem[\protect\citeauthoryear{Teslesca \& Lapenna}{2006}]{telesca2006}
Telesca, L., \& Lapenna V. 2006, Tecnophys., 423, 115

\bibitem[\protect\citeauthoryear{Taqqu et al.}{1995}]{taqqu}
Taqqu, M., Teverovsky, V., \& Willinger, W. 1995, Fractals, 3, 785

\bibitem[\protect\citeauthoryear{Watari}{1996}]{watari}
Watari, S. 1996. Solar Phys, 163, 371

\bibitem[\protect\citeauthoryear{Zhou}{2012}]{zhou}
Zhou, W. -X. 2012, Chaos, Solitons \& Fractals, 45, 147

\end{thebibliography}
\end{document}